\documentclass{svmono}
\usepackage[usenames]{color}
\usepackage[final]{graphicx}
\usepackage{framed}  
\usepackage{natbib}
\usepackage{amsmath,amsfonts,amssymb} 
\usepackage{bbm,alltt}
\usepackage{GCmacroBW}

\bibpunct{(}{)}{,}{a}{,}{,}
\renewcommand{\thesubsection}{$\;$}
\newcommand{\HM}{Metropolis-Hastings }
\newcommand{\MH}{Metropolis-Hastings}
\newcommand{\AR}{Accept-Reject}

\newcommand{\idxs}[1]{\relax}
\newcommand{\idxa}[1]{\relax}
\newcommand{\idxr}[1]{\relax}
\newcommand\odes{Odd-Numbered}

\begin{document}
\title{\LARGE Introducing Monte Carlo Methods with \R\\
{\Large Solutions to \odes~Exercises}
}
\author{Christian Robert\\Universit\'e Paris-Dauphine\\and\\George Casella\\University of Florida}

\maketitle

\frontmatter

\preface

\holmes{The scribes didn't have a large enough set from which to determine patterns.}{Brandon Sauderson}{The Hero of Ages}

\bigskip\noindent
This partial solution manual to our book {\em Introducing Monte Carlo Methods with R},
published by Springer Verlag in the {\sf User R!} series, on December 2009, has been compiled 
both from our own solutions and from homeworks 
written by the following Paris-Dauphine students in the 2009-2010 Master in Statistical Information Processing (TSI): 
Thomas Bredillet, Anne Sabourin, and Jiazi Tang. Whenever appropriate, the \R code
of those students has been identified by a \verb=# (C.) Name= in the text. 
We are grateful to those students for allowing us to use their solutions.
A few solutions in Chapter 4 are also taken {\em verbatim} from
the solution manual to {\em Monte Carlo Statistical Methods} compiled by Roberto Casarin from the University of Brescia
(and only available to instructors from Springer Verlag). 

We also incorporated in this manual indications about some typos found in the first printing that came to our
attention while composing this solution manual have been indicated as well.  Following the new ``print on demand"
strategy of Springer Verlag, these typos will not be found in the versions of the book purchased in the coming months and should
thus be ignored. (Christian Robert's book webpage at Universit\'e Paris-Dauphine \verb+www.ceremade.dauphine.fr/~xian/books.html+ 
is a better reference for the ``complete" list of typos.)

Reproducing the warning Jean-Michel Marin and Christian P.~Robert 
wrote at the start of the solution manual to {\em Bayesian Core}, let us stress here that
some self-study readers of {\em Introducing Monte Carlo Methods with {\sf R}} may come to the realisation that the solutions provided
here are too sketchy for them because the way we wrote those solutions assumes some minimal familiarity with the maths,
the probability theory and with the statistics behind the arguments. There is unfortunately a limit to the time and
to the efforts we can put in this solution manual and studying {\em Introducing Monte Carlo Methods with {\sf R}} 
requires some prerequisites in maths
(such as matrix algebra and Riemann integrals), in probability theory (such as the use of joint and conditional densities)
and some bases of statistics (such as the notions of inference, sufficiency and confidence sets) that we cannot cover here.
Casella and Berger (2001) is a good reference in case a reader is lost with the ``basic" concepts or sketchy math derivations.

We obviously welcome solutions, comments and questions on possibly erroneous or ambiguous solutions, as well as suggestions for
more elegant or more complete solutions: since this manual is distributed both freely and independently
from the book, it can be updated and corrected [almost] in real time! Note however that the {\sf R} codes given in the following
pages are not optimised because we prefer to use simple and understandable codes, rather than condensed and
efficient codes, both for time constraints and for pedagogical purposes: some codes were written by our students.
Therefore, if you find better [meaning, more efficient/faster] codes than those provided along those pages, we would be 
glad to hear from you, but that does not mean that we will automatically substitute your {\sf R} code for the current one,
because readability is also an important factor.

A final request: this manual comes in two versions, one corresponding to the odd-numbered exercises and 
freely available to everyone, and another one corresponding to a larger collection of exercises and with restricted access
to instructors only. Duplication and dissemination of the more extensive ``instructors only" version are obviously prohibited since,
if the solutions to most exercises become freely available, the appeal of using our book as a textbook will be severely
reduced. Therefore, if you happen to possess an extended version of the manual, please refrain from distributing
it and from reproducing it. 

\bigskip\noindent
{\bf Sceaux and Gainesville\hfil Christian P.~Robert~and~George Casella\break
\today\hfill}

\tableofcontents

\mainmatter
\setcounter{chapter}{0}
\chapter{Basic R programming}

\subsection{Exercise \ref{exo:baby}}

Self-explanatory.

\subsection{Exercise \ref{exo:helpme}}

Self-explanatory.

\subsection{Exercise \ref{exo:seq}}

One problem is the way in which \R handles parentheses.   So
\begin{verbatim}
> n=10
> 1:n
\end{verbatim}
produces 
\begin{verbatim}
1  2  3  4  5  6  7  8  9 10
\end{verbatim}
but
\begin{verbatim}
> n=10
> 1:n-1
\end{verbatim}
produces
\begin{verbatim}
0 1  2  3  4  5  6  7  8  9
\end{verbatim}
since the \verb+1:10+ command is executed first, then $1$ is subtracted.

The command \verb+seq(1,n-1,by=1)+ operates just as \verb+1:(n-1)+.
If $n$ is less than $1$ we can use something like \verb@seq(1,.05,by=-.01)@.  
Try it, and try some other variations.  

\subsection{Exercise \ref{pb:boot1}}

\begin{enumerate}
\renewcommand{\theenumi}{\alph{enumi}}
\item  To bootstrap the data you can use the code 
\begin{verbatim}
Boot=2500
B=array(0,dim=c(nBoot, 1))
for (i in 1:nBoot){
    ystar=sample(y,replace=T)
    B[i]=mean(ystar)
    }
\end{verbatim}
The quantile can be estimated with \verb+sort(B)[.95*nBoot]+, which in our case/sample is  $5.8478$.
\item  To get a confidence interval requires a double bootstrap.  That is, for each bootstrap sample we 
can get a point estimate of the $95\%$ quantile.  We can then run an histogram on these quantiles
with \verb@hist@, and get {\em their} upper and lower quantiles for a confidence region.
\begin{verbatim}
nBoot1=1000
nBoot2=1000
B1=array(0,dim=c(nBoot1, 1))
B2=array(0,dim=c(nBoot2, 1))
for (i in 1:nBoot1){
   ystar=sample(y,replace=T)
   for (j in 1:nBoot2)
      B2[j]=mean(sample(ystar,replace=T))
   B1[i]=sort(B2)[.95*nBoot2]
   }
\end{verbatim}
A $90\%$ confidence interval is given by
\begin{verbatim}
> c(sort(B1)[.05*nBoot1], sort(B1)[.95*nBoot1])
[1] 4.731 6.844
\end{verbatim}
or alternatively
\begin{verbatim}
> quantile(B1,c(.05,.95))
   5%    95%
4.731  6.844
\end{verbatim}
for the data in the book.  The command \verb@hist(B1)@ will give a histogram of the values.
\end{enumerate}

\subsection{Exercise \ref{exo:RvsC}}

If you type
\begin{verbatim}
> mean
function (x, ...)
UseMethod("mean")
<environment: namespace:base>
\end{verbatim}
you do not get any information about the function \verb+mean+ because it is not written in {\tt R}, while
\begin{verbatim}
> sd
function (x, na.rm = FALSE)
{
    if (is.matrix(x))
        apply(x, 2, sd, na.rm = na.rm)
    else if (is.vector(x))
        sqrt(var(x, na.rm = na.rm))
    else if (is.data.frame(x))
        sapply(x, sd, na.rm = na.rm)
    else sqrt(var(as.vector(x), na.rm = na.rm))
}
\end{verbatim}
shows \verb+sd+ is written in {\tt R}. The same applies to \verb+var+ and \verb+cov+.

\subsection{Exercise \ref{exo:attach}}

When looking at the description of \verb+attach+, you can see that this command allows to use
variables or functions that are in a database rather than in the current \verb=.RData=. Those
objects can be temporarily modified without altering their original format. (This is a fragile command
that we do not personaly recommend!)

The function \verb+assign+ is also rather fragile, but it allows for the creation and assignment of
an arbitrary number of objects, as in the documentation example:
\begin{verbatim}
for(i in 1:6) { #-- Create objects  'r.1', 'r.2', ... 'r.6' --
       nam <- paste("r",i, sep=".")
       assign(nam, 1:i)
      }
\end{verbatim}
which allows to manipulate the \verb+r.1+, \verb+r.2+, ..., variables.

\subsection{Exercise \ref{exo:dump&sink}}

This is mostly self-explanatory. If you type the help on each of those functions,
you will see examples on how they work. The most recommended \R function for saving
\R objects is \verb+save+. Note that, when using \verb\write\, the description states
\begin{verbatim}
The data (usually a matrix)  'x'  are written to file 
'file'. If  'x'  is a two-dimensional matrix you need 
to transpose it to get the columns in 'file' the same 
as those in the internal representation.
\end{verbatim}
Note also that \verb+dump+ and \verb+sink+ are fairly involved and should use with caution.

\subsection{Exercise \ref{exo:match}}

Take, for example {\tt a=3;x=c(1,2,3,4,5)} to see that they are the same, 
and, in fact, are the same as  \verb|max(which(x == a))|.   For 
\verb|y=c(3,4,5,6,7,8)|, try  \verb|match(x,y)| and \verb|match(y,x)| to 
see the difference.  In contrast, \verb|x%in%y| and \verb|y%in%y| return true/false tests.

\subsection{Exercise \ref{exo:timin}}

Running \verb=system.time= on the three sets of commands give
\begin{enumerate}
\item  0.004   0.000   0.071
\item 0       0       0
\item 0.000   0.000   0.001
\end{enumerate}
and the vectorial allocation is therefore the fastest\idxr{system.time@\verb+system.time+}.

\subsection{Exercise \ref{exo:unifix}}

The \R code is
\begin{verbatim}
> A=matrix(runif(4),ncol=2)
> A=A/apply(A,1,sum)
> apply(A%*%A,1,sum)
[1] 1 1
> B=A;for (t in 1:100) B=B%*%B
> apply(B,1,sum)
[1] Inf Inf
\end{verbatim}
and it shows that numerical inaccuracies in the product leads to the property to
fail when the power is high enough.

\subsection{Exercise \ref{exo:orange}}

The function \verb=xyplot= is part of the \verb+lattice+ library. Then
\begin{verbatim}
> xyplot(age ~ circumference, data=Orange)
> barchart(age ~ circumference, data=Orange)
> bwplot(age ~ circumference, data=Orange)
> dotplot(age ~ circumference, data=Orange)
\end{verbatim}
produce different representations of the dataset. Fitting a linear model is
simply done by \verb+lm(age ~ circumference, data=Orange)+
and using the tree index as an extra covariate leads to
\begin{verbatim}
>summary(lm(age ~ circumference+Tree, data=Orange))

Coefficients:
               Estimate Std. Error t value Pr(>|t|)
(Intercept)    -90.0596    55.5795  -1.620    0.116
circumference    8.7366     0.4354  20.066  < 2e-16 ***
Tree.L        -348.8982    54.9975  -6.344 6.23e-07 ***
Tree.Q         -22.0154    52.1881  -0.422    0.676
Tree.C          72.2267    52.3006   1.381    0.178
Tree^4          41.0233    52.2167   0.786    0.438
\end{verbatim}
meaning that only \verb=Tree.L= was significant.

\subsection{Exercise \ref{exo:sudoku}}

\begin{enumerate}                                                                                                                                    
\item A plain representation is
\begin{verbatim}
> s
      [,1] [,2] [,3] [,4] [,5] [,6] [,7] [,8] [,9]
 [1,]    0    0    0    0    0    6    0    4    0
 [2,]    2    7    9    0    0    0    0    5    0
 [3,]    0    5    0    8    0    0    0    0    2
 [4,]    0    0    2    6    0    0    0    0    0
 [5,]    0    0    0    0    0    0    0    0    0
 [6,]    0    0    1    0    9    0    6    7    3
 [7,]    8    0    5    2    0    0    4    0    0
 [8,]    3    0    0    0    0    0    0    8    5
 [9,]    6    0    0    0    0    0    9    0    1
\end{verbatim}
where empty slots are represented by zeros.

\item A simple cleaning of non-empty (i.e.~certain) slots is
\begin{verbatim}
for (i in 1:9)
for (j in 1:9){
  if (s[i,j]>0) pool[i,j,-s[i,j]]=FALSE
  }
\end{verbatim}

\item In {\tt R}, matrices (and arrays) are also considered as vectors. Hence \verb+s[i]+ represents
the $(1+\lfloor (i-1)/9 \rfloor,(i-1)\,\text{mod}\,9+1)$ entry of the grid.

\item This is self-explanatory. For instance,
\begin{verbatim}
> a=2;b=5
> boxa
[1] 1 2 3
> boxb
[1] 4 5 6
\end{verbatim}

\item The first loop checks whether or not, for each remaining possible integer, there exists
an identical entry in the same row, in the same column or in the same box. The second command
sets entries for which only one possible integer remains to this integer.

\item A plain \R program solving the grid is
\begin{verbatim}
while (sum(s==0)>0){
  for (i in sample(1:81)){
    if (s[i]==0){
       a=((i-1)%%9)+1
       b=trunc((i-1)/9)+1
       boxa=3*trunc((a-1)/3)+1
       boxa=boxa:(boxa+2)
       boxb=3*trunc((b-1)/3)+1
       boxb=boxb:(boxb+2)

       for (u in (1:9)[pool[a,b,]]){
         pool[a,b,u]=(sum(u==s[a,])+sum(u==s[,b])
                +sum(u==s[boxa,boxb]))==0
         }

       if (sum(pool[a,b,])==1){ 
         s[i]=(1:9)[pool[a,b,]]
         }

       if (sum(pool[a,b,])==0){
          print("wrong sudoku")
          break()
          }
       }
    }
  }
\end{verbatim}
and it stops with the outcome
\begin{verbatim}
> s
      [,1] [,2] [,3] [,4] [,5] [,6] [,7] [,8] [,9]
 [1,]    1    3    8    5    2    6    7    4    9
 [2,]    2    7    9    3    4    1    8    5    6
 [3,]    4    5    6    8    7    9    3    1    2
 [4,]    7    4    2    6    3    5    1    9    8
 [5,]    9    6    3    1    8    7    5    2    4
 [6,]    5    8    1    4    9    2    6    7    3
 [7,]    8    9    5    2    1    3    4    6    7
 [8,]    3    1    7    9    6    4    2    8    5
 [9,]    6    2    4    7    5    8    9    3    1
\end{verbatim}
which is the solved Sudoku.
\end{enumerate}

\chapter{Random Variable Generation}
\renewcommand\AR{Accept--Reject~}

\subsection{Exercise \ref{pb:discretePIT}}

For a random variable $X$ with cdf $F$, if
\[
F^{-}(u)=\inf\{ x,F(x)\leq u\},
\] 
then, for $U\sim\mathcal{U}[0,1]$, for all $y \in \mathbb{R}$,
\begin{eqnarray*}
\mathbb{P}(F^{-}(U)\leq y)&=&\mathbb{P}(\inf\{ x,F(x)\leq U\}\leq y)\\
 && =\mathbb{P}(F(y)\geq U)\qquad\textrm{ as $F$ is non-decreasing }\\
 && =F(y)\qquad\qquad\textrm{ as $U$ is uniform}
\end{eqnarray*}

\subsection{Exercise \ref{pb:boxmuller}}

\begin{enumerate}
\renewcommand{\theenumi}{\alph{enumi}}
\item It is easy to see that $\BE[U_1]=0$, and a standard calculation shows that $\text{var}(U_1)= 1/12$, from which the result follows.
\item  Histograms show that the tails of the $12$ uniforms are not long enough.  Consider the code
\begin{verbatim}
nsim=10000
u1=runif(nsim)
u2=runif(nsim)
X1=sqrt(-2*log(u1))*cos(2*pi*u2)
X2=sqrt(-2*log(u1))*sin(2*pi*u2)
U=array(0,dim=c(nsim,1))
for(i in 1:nsim)U[i]=sum(runif(12,-.5,.5))
par(mfrow=c(1,2))
hist(X1)
hist(U)
a=3
mean(X1>a)
mean(U>a)
mean(rnorm(nsim)>a)
1-pnorm(a)
\end{verbatim}

\item You should see the difference in the tails of the histogram.  Also, the numerical output from the above is
\begin{verbatim}
[1] 0.0016
[1] 5e-04
[1] 0.0013
[1] 0.001349898
\end{verbatim}
where we see that the Box-Muller and \verb+rnorm+ are very good when compared with the exact \verb+pnorm+.  
Try this calculation for a range of \verb+nsim+ and \verb+a+.
\end{enumerate}

\subsection{Exercise \ref{exo:acceP}}

For $U\sim\mathcal{U}_{[0,1]}$, $Y\sim g(y)$, and $X\sim f(x)$, such that
$f/g\leq M$, the acceptance condition in the Accept--Reject algorithm is that $U\leq f(Y)/(Mg(Y)).$
The probability of acceptance is thus
\begin{align*}
\mathbb{P}(U  \leq  f(Y)\big/ Mg(Y))&=\int_{-\infty}^{+\infty}\int_{0}^{\frac{f(y}{Mg(y)}}\,\text{d}ug(y)\,\text{d}y\\
   & =\int_{-\infty}^{+\infty}\frac{f(y)}{Mg(y)}g(y)\,\text{d}y\\
   & =\frac{1}{M}\int_{-\infty}^{+\infty}f(y)\,\text{d}y\\
   & =\frac{1}{M}\,.
\end{align*}
Assume $f/g$ is only known up to a normalising constant, i.e.
$f/g=k.\tilde{f}/\tilde{g}$, with $\tilde{f}/\tilde{g}\leq\tilde{M}$,
$\tilde{M}$ being a well-defined upper bound different from $M$ because of the missing
normalising constants. Since $Y\sim g$,
\begin{align*}
\mathbb{P}(U \leq  \tilde{f}(Y)\big/ \tilde{M}\tilde{g}(Y))
 & =\int_{-\infty}^{+\infty}\int_{0}^{\frac{\tilde{f}(y}{\tilde{M}\tilde{g}(y)}}\,\text{d}ug(y)\,\text{d}y\\
 & =\int_{-\infty}^{+\infty}\frac{\tilde{f}(y)}{\tilde{M}\tilde{g}(y)}g(y)\,\text{d}y\\
 & =\int_{-\infty}^{+\infty}\frac{f(y)}{k\tilde{M}g(y)}g(y)\,\text{d}y\\
 & =\frac{1}{k\tilde{M}}\,.
\end{align*}
Therefore the missing constant is given by
$$
k=1\bigg/ \tilde{M.}\mathbb{P}(U\leq\tilde{f}(Y)\big/ \tilde{M}\tilde{g}(Y))\,,
$$
which can be estimated from the empirical acceptance rate.

\subsection{Exercise \ref{exo:trueMax}}

The ratio is equal to
$$
\frac{\Gamma(\alpha+\beta)}{\Gamma(\alpha)\Gamma(\beta)}\,
\frac{\Gamma(a)\Gamma(b)}{\Gamma(a+b)}\, x^{\alpha-a}\,(1-x)^{\beta-b}
$$
and it will not diverge at $x=0$ only if $a\le \alpha$ and at $x=1$ only if $b\le \beta$.
The maximum is attained for
$$
\frac{\alpha-a}{x^\star} = \frac{\beta-b}{1-x^\star}\,,
$$
i.e.~is
$$
M_{a,b}=\frac{\Gamma(\alpha+\beta)}{\Gamma(\alpha)\Gamma(\beta)}\,
\frac{\Gamma(a)\Gamma(b)}{\Gamma(a+b)}\, \frac{(\alpha-a)^{\alpha-a}(\beta-b)^{\beta-b}}
{(\alpha-a+\beta-b)^{\alpha-a+\beta-b}}\,.
$$
The analytic study of this quantity as a function of $(a,b)$ is quite delicate but if we define
\begin{verbatim}
mab=function(a,b){
  lgamma(a)+lgamma(b)+(alph-a)*log(alph-a)+(beta-b)*log(beta-b)
  -(alph+bet-a-b)*log(alph+bet-a-b)}
\end{verbatim}
it is easy to see using \verb=contour= on a sequence of $a$'s and $b$'s that the maximum of
$M_{a,b}$ is achieved over integer values when $a=\lfloor \alpha \rfloor$ and
$b=\lfloor \beta \rfloor$.

\subsection{Exercise \ref{exo:inzabove}}

Given $\theta$, exiting the loop is driven by $X=x_0$, which indeed has a probability
$f(x_0|\theta)$ to occur. If $X$ is a discrete random variable, this is truly a probability, while,
if $X$ is a continuous random variable, this is zero. The distribution of the exiting $\theta$ is
then dependent on the event $X=x_0$ taking place, i.e.~is proportional to $\pi(\theta)f(x_0|\theta)$,
which is exactly $\pi(\theta|x_0)$.

\subsection{Exercise \ref{pb:hist}}

\begin{enumerate}
\renewcommand{\theenumi}{\alph{enumi}}
\item Try the \R code
\begin{verbatim}
nsim<-5000
n=25;p=.2;
cp=pbinom(c(0:n),n,p)
X=array(0,c(nsim,1))
for(i in 1:nsim){
   u=runif(1)
   X[i]=sum(cp<u)
   }
hist(X,freq=F)
lines(1:n,dbinom(1:n,n,p),lwd=2)
\end{verbatim}
which produces a histogram and a mass function for the binomial $\mathcal{B}(25,.2)$.

To check timing, create the function
\begin{verbatim}
MYbinom<-function(s0,n0,p0){ 
  cp=pbinom(c(0:n0),n0,p0)
  X=array(0,c(s0,1))
  for (i in 1:s0){
        u=runif(1)
        X[i]=sum(cp<u)
        }
  return(X)
  }
\end{verbatim}
and use \verb+system.time(rbinom(5000,25,.2))+ and \verb+system.time(MYbinom(5000,25,.2))+ 
to see how much faster \R is.

\item  Create the \R functions {\tt Wait} and {\tt Trans}:
\begin{verbatim}
Wait<-function(s0,alpha){
  U=array(0,c(s0,1))
  for (i in 1:s0){
     u=runif(1)
     while (u > alpha) u=runif(1)
     U[i]=u
     }
  return(U)
  }

Trans<-function(s0,alpha){
  U=array(0,c(s0,1))
  for (i in 1:s0) U[i]=alpha*runif(1)
  return(U)
  }
\end{verbatim}
Use \verb+hist(Wait(1000,.5))+ and \verb+hist(Trans(1000,.5))+ to see the
corresponding histograms.  Vary $n$ and $\alpha$.  Use the \verb+system.time+ command as in part a to see the timing.  
In particular, \verb+Wait+ is very bad if $\alpha$ is small.
\end{enumerate}

\subsection{Exercise \ref{pb:pareto_gen}}

The cdf of the Pareto $\CP(\alpha)$ distribution is
$$
F(x)=1-x^{-\alpha}
$$
over $(1,\infty)$. Therefore, $F^{-1}(U)=(1-U)^{-1/\alpha}$, which is also
the $-1/\alpha$ power of a uniform variate.

\subsection{Exercise \ref{pb:specific}}

Define the \R functions
\begin{verbatim}
  Pois1<-function(s0,lam0){
     spread=3*sqrt(lam0)
     t=round(seq(max(0,lam0-spread),lam0+spread,1))
     prob=ppois(t,lam0)
     X=rep(0,s0)
     for (i in 1:s0){
        u=runif(1)
        X[i]=max(t[1],0)+sum(prob<u)-1
        }
     return(X)
     }
\end{verbatim}
and 
\begin{verbatim}
  Pois2<-function(s0,lam0){
     X=rep(0,s0)
     for (i in 1:s0){
     sum=0;k=1
     sum=sum+rexp(1,lam0)
     while (sum<1){ sum=sum+rexp(1,lam0);k=k+1}
     X[i]=k
     }
     return(X)
     }
\end{verbatim}
and then run the commands
\begin{verbatim}
> nsim=100
> lambda=3.4
> system.time(Pois1(nsim,lambda))
   user  system elapsed
  0.004   0.000   0.005
> system.time(Pois2(nsim,lambda))
   user  system elapsed
  0.004   0.000   0.004
> system.time(rpois(nsim,lambda))
   user  system elapsed
      0       0       0
\end{verbatim}
for other values of \verb+nsim+ and \verb+lambda+.  You will see that \verb@rpois@ is by far the best, with the exponential generator 
(\verb#Pois2#) not being very good for large $\lambda$'s.  Note also that \verb@Pois1@ is not appropriate for small $\lambda$'s since 
it could then return negative values.

\subsection{Exercise \ref{pb:gammaAR}}

\begin{enumerate}
\renewcommand{\theenumi}{\alph{enumi}}
\item Since, if $X\sim \mathcal{G}a(\alpha,\beta)$, then $\beta X=\sum_{j=1}^{\alpha} \beta X_{j} \sim \mathcal{G}a(\alpha,1)$,
$\beta$ is the inverse of a scale parameter.
\item The Accept-Reject ratio is given by
$$
\dfrac{f(x)}{g(x)} \propto \dfrac{x^{n-1}\,e^{-x}}{\lambda\,e^{-\lambda x}}=\lambda^{-1} x^{n-1} e^{-(1-\lambda)x}\,.
$$
The maximum of this ratio is obtained for 
$$
\dfrac{n-1}{x^\star} - (1-\lambda) = 0\,,\quad\text{i.e. for}\quad x^\star = \dfrac{n-1}{1-\lambda}\,.
$$
Therefore, 
$$
M\propto \lambda^{-1} \left( \dfrac{n-1}{1-\lambda} \right)^{n-1} \,e^{-(n-1)}
$$
and this upper bound is minimised in $\lambda$ when $\lambda=1/n$.

\item If $g$ is the density of the $\mathcal{G}a(a,b)$ distribution and $f$ the
density of the $\mathcal{G}a(\alpha,1)$ distribution, 
$$
g(x) = \frac{x^{a-1}e^{-bx}b^{a}}{\Gamma(a)} \quad\text{and}\quad f(x) = \frac{x^{\alpha-1}e^{-x}}{\Gamma(\alpha)} 
$$
the Accept-Reject ratio is given by
$$ 
\dfrac{f(x)}{g(x)} = \dfrac{x^{\alpha-1}e^{-x} \Gamma(a)}{\Gamma(\alpha) b^{a} x^{a-1}e^{-bx}} \propto
b^{-a}x^{\alpha-a}e^{-x(1-b)} \,.
$$
Therefore,
$$
\dfrac{\partial}{\partial x} \dfrac{f}{g}  = b^{a} e^{-x(1-b)}x^{\alpha-a-1}\left\{(\alpha-a)-(1-b)x\right\} 
$$
provides $x^\star = {\alpha-a}\big/{1-b} $ as the argument of the maximum of the ratio, since $\frac{f}{g} (0)= 0$.
The upper bound $M$ is thus given by
$$
M(a,b)=b^{-a}\left(\dfrac{\alpha-a}{1-b}\right)^{\alpha-a}e^{-\left(\frac{\alpha-a}{1-b}\right)*(1-b)} 
      =b^{-a}\left(\frac{\alpha-a}{(1-b) e}\right)^{\alpha-a}\,. 
$$
It obviously requires $b<1$ and $a<\alpha$.

\item {\bf Warning: there is a typo in the text of the first printing, it should be:}
\begin{rema}
Show that the maximum of $b^{-a}(1 - b)^{a-\alpha}$ is attained at $b = a/\alpha$, and hence the optimal choice of $b$
for simulating ${\cal{G}}a(\alpha,1)$ is $b=a/\alpha$, which gives the same mean for both ${\cal{G}}a(\alpha,1)$ and ${\cal{G}}a(a,b)$.
\end{rema}
With this modification, the maximum of $M(a,b)$ in $b$ is obtained by derivation, i.e.~for $b$ solution of
$$
\dfrac{a}{b}-\dfrac{\alpha-a}{1-b}=0\,,
$$
which leads to $b = a/\alpha$ as the optimal choice of $b$. Both ${\cal{G}}a(\alpha,1)$ and ${\cal{G}}a(a,a/\alpha)$ have the same mean $\alpha$.

\item Since
$$
M(a,a/\alpha) = (a/\alpha)^{-a}\left(\frac{\alpha-a}{(1-a/\alpha) e}\right)^{\alpha-a}
              = (a/\alpha)^{-a} \alpha^{\alpha-a} = \alpha^\alpha/a^a,,
$$
$M$ is decreasing in $a$ and the largest possible value is indeed $a=\lfloor \alpha \rfloor$.
\end{enumerate}

\subsection{Exercise \ref{pb:Norm-DEAR}}

The ratio $f/g$ is
$$
\dfrac{f(x)}{g(x)} = \dfrac{\exp\{-x^2/2\}/\sqrt{2\pi}}{\alpha\exp\{-\alpha|x|\}/2}
=\dfrac{\sqrt{2/\pi}}{\alpha}\,\exp\{\alpha|x|-x^2/2\}
$$
and it is maximal when $x=\pm\alpha$, so $M=\sqrt{2/\pi}\exp\{\alpha^2/2\}/\alpha$.
Taking the derivative in $\alpha$ leads to the equation
$$
\alpha-\frac{1}{\alpha^2} =0\,,
$$
that is, indeed, to $\alpha=1$.

\subsection{Exercise \ref{pb:noncen_chi}}

{\bf Warning: There is a typo in this exercise, it should be:}
\begin{rema}
\begin{enumerate}
\renewcommand{\theenumi}{(\roman{enumi})}
\item a mixture representation (\ref{eq:mixture_def}), where
$g(x \vert y)$ is the density of $\chi_{p+2y}^{2}$ and $p(y)$ is the density of $\CP(\lambda/2)$, and
\item the sum of a $\chi_{p-1}^{2}$ random variable and the square of a ${\cal N}(\sqrt{\lambda},1)$.
\end{enumerate}
\begin{enumerate}
\renewcommand{\theenumi}{\alph{enumi}}
\item Show that both those representations hold.
\item Compare the corresponding algorithms that can be derived from these
representations among themselves and also with {\tt rchisq} for small and large values of $\lambda$.
\end{enumerate}
\end{rema}
If we use the definition of the noncentral chi squared distribution, $\chi_{p}^{2}(\lambda)$ as corresponding to
the distribution of the squared norm $||x||^2$ of a normal vector $x\sim\mathcal{N}_p(\theta,I_p)$
when $\lambda=||\theta||^2$, this distribution is invariant by rotation over the normal vector and it is therefore
the same as when $x\sim\mathcal{N}_p((0,\ldots,0,\sqrt{\lambda}),I_p)$, hence leading to the representation (ii),
i.e.~as a sum of a $\chi_{p-1}^{2}$ random variable and of the square of a ${\cal N}(||\theta||,1)$ variable.
Representation (i) holds by a regular mathematical argument based on the series expansion of the modified
Bessel function since the density of a non-central chi-squared distribution is
$$
f(x|\lambda) = {1\over 2} (x/\lambda)^{(p-2)/4} I_{(p-2)/2}(\sqrt{\lambda x}) e^{-(\lambda+x)/2}\,,
$$
where
$$
I_\nu (t) = \left({t\over 2}\right)^{\nu} \sum_{k=0}^\infty
{(z/2)^{2k} \over k! \Gamma(\nu+k+1)}.
$$

Since \verb=rchisq= includes an optional non-centrality parameter \verb=nc=, it can be used
to simulate directly a noncentral chi-squared distribution. The two scenarios (i) and (ii) lead
to the following \R codes.
\begin{verbatim}
> system.time({x=rchisq(10^6,df=5,ncp=3)})
   user  system elapsed
> system.time({x=rchisq(10^6,df=4)+rnorm(10^6,mean=sqrt(3))^2})
   user  system elapsed
  1.700   0.056   1.757
> system.time({x=rchisq(10^6,df=5+2*rpois(10^6,3/2))})
   user  system elapsed
  1.168   0.048   1.221
\end{verbatim}
Repeated experiments with other values of $p$ and $\lambda$ lead to the same conclusion that the
Poisson mixture representation is the fastest.\\

\subsection{Exercise \ref{pb:bayesAR}}

Since the ratio $\pi(\theta|{\mathbf x})/\pi(\theta)$ is the likelihood, it is obvious that
the optimal bound $M$ is the likelihood function evaluated at the MLE (assuming $\pi$ is a true
density and not an improper prior).

Simulating from the posterior can then be done via
\begin{verbatim}
theta0=3;n=100;N=10^4
x=rnorm(n)+theta0
lik=function(the){prod(dnorm(x,mean=the))}
M=optimise(f=function(the){prod(dnorm(x,mean=the))},
  int=range(x),max=T)$obj
theta=rcauchy(N)
res=(M*runif(N)>apply(as.matrix(theta),1,lik));print(sum(res)/N)
while (sum(res)>0){le=sum(res);theta[res]=rcauchy(le)
res[res]=(M*runif(le)>apply(as.matrix(theta[res]),1,lik))}
\end{verbatim}
The rejection rate is given by $0.9785$, which means that the Cauchy proposal is quite inefficient.
An empirical confidence (or credible) interval at the level $95\%$ on $\theta$ is $(2.73,3.799)$.
Repeating the experiment with $n=100$ leads (after a while) to the interval $(2.994,3.321)$, there is
therefore an improvement.

\chapter{Monte Carlo Integration}

\subsection{Exercise \ref{pb:norm_cauchy}}

\begin{enumerate}
\item The plot of the integrands follows from a simple \R program:
\begin{verbatim}
f1=function(t){  t/(1+t*t)*exp(-(x-t)^2/2)}
f2=function(t){  1/(1+t*t)*exp(-(x-t)^2/2)}
plot(f1,-3,3,col=1,ylim=c(-0.5,1),xlab="t",ylab="",ty="l")
plot(f2,-3,3,add=TRUE,col=2,ty="l")
legend("topright", c("f1=t.f2","f2"), lty=1,col=1 :2)
\end{verbatim}
Both numerator and denominator are expectations under the Cauchy distribution. They can therefore
be approximated directly by
\begin{verbatim}
Niter=10^4
co=rcauchy(Niter)
I=mean(co*dnorm(co,mean=x))/mean(dnorm(co,mean=x))
\end{verbatim}
We thus get
\begin{verbatim}
> x=0
> mean(co*dnorm(co,mean=x))/mean(dnorm(co,mean=x))
[1] 0.01724
> x=2
> mean(co*dnorm(co,mean=x))/mean(dnorm(co,mean=x))
[1] 1.295652
> x=4
> mean(co*dnorm(co,mean=x))/mean(dnorm(co,mean=x))
[1] 3.107256
\end{verbatim}
\item Plotting the convergence of those integrands can be done via
\begin{verbatim}
# (C.) Anne Sabourin, 2009
x1=dnorm(co,mean=x)
estint2=cumsum(x1)/(1:Niter)
esterr2=sqrt(cumsum((x1-estint2)^2))/(1:Niter)
x1=co*x1
estint1=cumsum(x1))/(1:Niter)
esterr2=sqrt(cumsum((x1-estint1)^2))/(1:Niter)
par(mfrow=c(1,2))
plot(estint1,type="l",xlab="iteration",ylab="",col="gold")
lines(estint1-2*esterr1,lty=2,lwd=2)
lines(estint1+2*esterr1,lty=2,lwd=2)
plot(estint2,type="l",xlab="iteration",ylab="",col="gold")
lines(estint2-2*esterr2,lty=2,lwd=2)
lines(estint2+2*esterr2,lty=2,lwd=2)
\end{verbatim}
Because we have not yet discussed the evaluation of the error for a ratio of estimators, we consider
both terms of the ratio separately. The empirical variances $\hat\sigma$ are given by \verb+var(co*dnorm(co,m=x))+
and \verb+var(dnorm(co,m=x))+ and solving $2\hat\sigma/\sqrt{n}<10^{-3}$ leads to an evaluation of the number of
simulations necessary to get $3$ digits of accuracy.
\begin{verbatim}
> x=0;max(4*var(dnorm(co,m=x))*10^6,
+ 4*var(co*dnorm(co,m=x))*10^6)
[1] 97182.02
> x=2; 4*10^6*max(var(dnorm(co,m=x)),var(co*dnorm(co,m=x)))
[1] 220778.1
> x=4; 10^6*4*max(var(dnorm(co,m=x)),var(co*dnorm(co,m=x)))
[1] 306877.9
\end{verbatim}
\item A similar implementation applies for the normal simulation, replacing \verb=dnorm= with \verb=dcauchy= in the
above. The comparison is clear in that the required number of normal simulations when $x=4$ is $1398.22$, to compare
with the above $306878$.
\end{enumerate}

\subsection{Exercise \ref{exo:tailotwo}}

Due to the identity  
$$
\mathbb{P}(X>20) = \int_{20}^{\infty}\dfrac{\exp(-\frac{x^2}{2})}{\sqrt{2\pi}}\text{d}x 
= \int_{0}^{1/20}\frac{\exp(-\frac{1}{2*u^2})}{20 u^2 \sqrt{2\pi}}20 \text{d}u\,, 
$$
we can see this integral as an expectation under the $\mathcal{U}(0,1/20)$
distribution and thus use a Monte Carlo approximation to $\mathbb{P}(X>20)$.
The following \R code monitors the convergence of the corresponding approximation.
\begin{verbatim}
# (C.) Thomas Bredillet, 2009
h=function(x){ 1/(x^2*sqrt(2*pi)*exp(1/(2*x^2)))}
par(mfrow=c(2,1))
curve(h,from=0,to=1/20,xlab="x",ylab="h(x)",lwd="2")
I=1/20*h(runif(10^4)/20)
estint=cumsum(I)/(1:10^4)
esterr=sqrt(cumsum((I-estint)^2))/(1:10^4)
plot(estint,xlab="Iterations",ty="l",lwd=2,
ylim=mean(I)+20*c(-esterr[10^4],esterr[10^4]),ylab="")
lines(estint+2*esterr,col="gold",lwd=2)
lines(estint-2*esterr,col="gold",lwd=2)
\end{verbatim}
The estimated probability is $2.505 e-89$ with an error of $\pm 3.61 e-90$, compared with 
\begin{verbatim}
> integrate(h,0,1/20)
2.759158e-89 with absolute error < 5.4e-89
> pnorm(-20)
[1] 2.753624e-89
\end{verbatim}

\subsection{Exercise \ref{exo:fperron+}}

{\bf Warning: due to the (late) inclusion of an extra-exercise in the book, 
the ``above exercise" actually means Exercise \ref{exo:tailotwo}!!!}\\

\begin{figure}
\begin{center}
\centerline{\includegraphics[width=.95\textwidth]{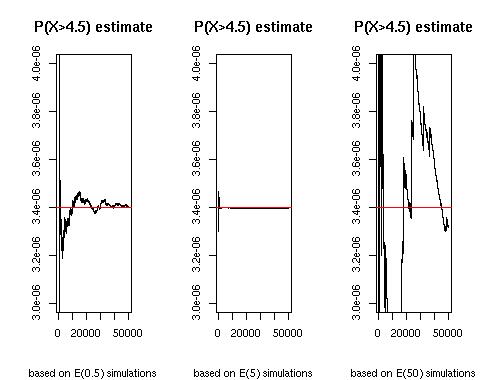}}
\caption{\label{fig:nortail}
Comparison of three importance sampling approximations to the normal tail probability $\mathbb{P}(Z>4.5)$ based
on a truncated $\mathcal{E}xp(\lambda)$ distribution with $\lambda=.5,5.50$. The straight red line is the true value.}
\end{center}
\end{figure}

When $Z\sim\mathcal{N}(0,1)$, with density $f$, the quantity of interest is $\mathbb{P}(Z>4.5)$, 
i.e.~$\mathbb{E}^{f}[\mathbb{I}_{Z>4.5}]$. When $g$ is the density of
the exponential $\mathcal{E}xp(\lambda)$ distribution truncated at $4.5$,
$$
g(y)=\frac{1_{y>4.5}\lambda\exp(-\lambda y)}{\int_{-4.5}^{\infty}\lambda\exp(-\lambda y)\,\text{d}y}
=\lambda e^{-\lambda(y-4.5)}\mathbb{I}_{y>4.5}\,,
$$
simulating iid $Y^{(i)}$'s from $g$ is straightforward. Given that the indicator function
$\mathbb{I}_{Y>4.5}$ is then always equal to $1$, $\mathbb{P}(Z>4.5)$ is estimated by
$$
\hat{h}_{n}=\frac{1}{n}\sum_{i=1}^{n}\frac{f(Y^{(i)})}{g(Y^{(i)})}.
$$
A corresponding estimator of its variance is
$$
v_{n}=\frac{1}{n²}\sum_{i=1}^{n}(1-\hat{h}_{n})^{2}{f(Y^{(i)})}\big/{g(Y^{(i)})}\,.
$$
The following \R code monitors the convergence of the estimator (with $\lambda=1,10$)
\begin{verbatim}
# (C.) Anne Sabourin, 2009
Nsim=5*10^4
x=rexp(Nsim)
par(mfcol=c(1,3))
for (la in c(.5,5,50)){
  y=(x/la)+4.5
  weit=dnorm(y)/dexp(y-4.5,la)
  est=cumsum(weit)/(1:Nsim)
  varest=cumsum((1-est)^2*weit/(1:Nsim)^2)
  plot(est,type="l",ylim=c(3e-6,4e-6),main="P(X>4.5) estimate",
  sub=paste("based on E(",la,") simulations",sep=""),xlab="",ylab="")
  abline(a=pnorm(-4.5),b=0,col="red")
  }
\end{verbatim}
When evaluating the impact of $\lambda$ on the variance (and hence on the convergence) of the estimator,
similar graphs can be plotted for different values of $\lambda$. This experiment does not exhibit a clear
pattern, even though large values of $\lambda$, like $\lambda=20$ appear to slow down convergence very much.
Figure \ref{fig:nortail} shows the output of such a comparison. Picking $\lambda=5$ seems however to produce
a very stable approximation of the tail probability.

\subsection{Exercise \ref{pb:some_jump}}

While the expectation of $\sqrt{x/(1-x)}$ is well defined for $\nu>1/2$, the integral of
$x/(1-x)$ against the $t$ density does not exist for any $\nu$. Using an importance sampling representation,
$$
\int \frac{x}{1-x}\,\frac{f^2(x)}{g(x)}\,\text{d}x = \infty
$$
if $g(1)$ is finite. The integral will be finite around $1$ when $1/(1-t)g(t)$ is integrable, which
means that $g(t)$ can go to infinity at any rate. For instance, if $g(t)\approx(1-t)^{-\alpha}$ around
$1$, any $\alpha>0$ is acceptable.\\

\subsection{Exercise \ref{pb:bayesAR2}}

As in Exercise \ref{pb:norm_cauchy}, 
the quantity of interest is $\delta^{\pi}(x)=\mathbb{E}^{\pi}(\theta|x)=\int\theta\pi(\theta|x)\,\text{d}\theta$
where $x\sim\mathcal{N}(\theta,1)$ and $\theta\sim\mathcal{C}(0,1)$. The target
distribution is
$$
\pi(\theta|x)\propto{\pi(\theta)e^{-(x-\theta)^{2}/2}} = f_{x}(\theta)\,.
$$
A possible importance function is the prior distribution, $$g(\theta)=\frac{1}{\pi(1+\theta^{2})}$$
and for every $\theta\in\mathbb{R}$, $\frac{f_{x}(\theta)}{g(\theta)}\leq M$, when $M=\pi$.
Therefore, generating from the prior $g$ and accepting simulations according to the
Accept-Reject ratio provides a sample from 
$\pi(\theta|x)$. The empirical mean of this sample is then a converging estimator of $\mathbb{E}^{\pi}(\theta|x).$
Furthermore, we directly deduce the estimation error for $\delta$.
A graphical evaluation of the convergence is given by the following \R program:
\begin{verbatim}
f=function(t){ exp(-(t-3)^2/2)/(1+t^2)}
M=pi
Nsim=2500
postdist=rep(0,Nsim)
for (i in 1:Nsim){
   u=runif(1)*M
   postdist[i]=rcauchy(1)
   while(u>f(postdist[i])/dcauchy(postdist[i])){
     u=runif(1)*M
     postdist[i]=rcauchy(1)
     }}
estdelta=cumsum(postdist)/(1:Nsim)
esterrd=sqrt(cumsum((postdist-estdelta)^2))/(1:Nsim)
par(mfrow=c(1,2))
C1=matrix(c(estdelta,estdelta+2*esterrd,estdelta-2*esterrd),ncol=3)
matplot(C1,ylim=c(1.5,3),type="l",xlab="Iterations",ylab="")
plot(esterrd,type="l",xlab="Iterations",ylab="")
\end{verbatim}

\subsection{Exercise \ref{pb:smalltail}}

\begin{enumerate}
\renewcommand{\theenumi}{\alph{enumi}}
\item If $X \sim \mathcal{E}xp(1)$ then for $x \ge a$,   
$$
\mathbb{P}[a + X < x] = \int_{0}^{x-a} \exp(-t)\,\text{d}t 
= \int_{a}^{x} \exp(-t+a)\,\text{d}t = \mathbb{P}(Y < x) 
$$
when $Y \sim\mathcal{E}xp^{+}(a,1)$, 

\item If $ X \sim \chi^{2}_{3}$, then
\begin{align*}
\mathbb{P}(X>25) 
&= \int_{25}^{+\infty} \frac{2^{-3/2}}{\Gamma(\frac{3}{2})}\,x^{1/2}\exp(-x/2)\,\text{d}x\\ 
&= \int_{12.5}^{+\infty} \frac{\sqrt(x)\exp(-12.5)}{\Gamma(\frac{3}{2})}\exp(-x+12.5)\,\text{d}x\,. 
\end{align*}
The corresponding \R code
\begin{verbatim} 
# (C.) Thomas Bredilllet, 2009
h=function(x){ exp(-x)*sqrt(x)/gamma(3/2)}
X = rexp(10^4,1) + 12.5
I=exp(-12.5)*sqrt(X)/gamma(3/2)
estint=cumsum(I)/(1:10^4)
esterr=sqrt(cumsum((I-estint)^2))/(1:10^4)
plot(estint,xlab="Iterations",ty="l",lwd=2,
ylim=mean(I)+20*c(-esterr[10^4],esterr[10^4]),ylab="")
lines(estint+2*esterr,col="gold",lwd=2)
lines(estint-2*esterr,col="gold",lwd=2)
\end{verbatim}
gives an evaluation of the probability as $1.543e-05 $ with a $10^{-8}$ error, to compare
with
\begin{verbatim} 
> integrate(h,12.5,Inf)
1.544033e-05 with absolute error < 3.4e-06
> pchisq(25,3,low=F)
[1] 1.544050e-05
\end{verbatim}

Similarly, when $X \sim t_{5} $, then
$$
\mathbb{P}(X>50) = \int_{50}^{\infty} \dfrac{\Gamma(3)}{\sqrt(5*\pi)\Gamma(2,5)
(1+\frac{t^2}{5})^{3}\exp(-t+50)}\exp(-t+50)\,\text{d}t 
$$
and a corresponding \R code
\begin{verbatim}
# (C.) Thomas Bredilllet, 2009
h=function(x){ 1/sqrt(5*pi)*gamma(3)/gamma(2.5)*1/(1+x^2/5)^3}
integrate(h,50,Inf)
X = rexp(10^4,1) + 50
I=1/sqrt(5*pi)*gamma(3)/gamma(2.5)*1/(1+X^2/5)^3*1/exp(-X+50)
estint=cumsum(I)/(1:10^4)
esterr=sqrt(cumsum((I-estint)^2))/(1:10^4)
plot(estint,xlab="Mean and error range",type="l",lwd=2,
ylim=mean(I)+20*c(-esterr[10^4],esterr[10^4]),ylab="")
lines(estint+2*esterr,col="gold",lwd=2)
lines(estint-2*esterr,col="gold",lwd=2)
\end{verbatim}
As seen on the graph, this method induces jumps in the convergence patterns. Those jumps are indicative of
variance problems, as should be since the estimator does not have a finite variance in this case. The value
returned by this approach differs from alternatives evaluations:
\begin{verbatim}
>  mean(I)
[1] 1.529655e-08
> sd(I)/10^2
[1] 9.328338e-10
> integrate(h,50,Inf)
3.023564e-08 with absolute error < 2e-08
> pt(50,5,low=F)
[1] 3.023879e-08
\end{verbatim}
and cannot be trusted.

\item {\bf Warning: There is a missing line in the text of this question, which should read:}
\begin{rema}
\noindent Explore the gain in efficiency from this method.  Take $a=4.5$ in part (a) and run an
experiment to determine how many normal $\mathcal{N}(0,1)$ random variables would be needed to calculate $P(Z > 4.5)$
to the same accuracy obtained from using $100$ random variables in this importance sampler.
\end{rema}

If we use the representation
$$
\mathbb{P}(Z>4.5) = \int_{4.5}^\infty \varphi(z)\,\text{d}z = \int_0^\infty \varphi(x+4.5)
\exp(x)\exp(-x)\,\text{d}x\,,
$$
the approximation based on $100$ realisations from an $\mathcal{E}xp(1)$ distribution, $x_1m\ldots,x_100$, is
$$
\frac{1}{100}\,\sum_{i=1}^{100} \varphi(x_i+4.5) \exp(x_i)
$$
and the \R code
\begin{verbatim}
> x=rexp(100)
> mean(dnorm(x+4.5)*exp(x))
[1] 2.817864e-06
> var(dnorm(x+4.5)*exp(x))/100
[1] 1.544983e-13
\end{verbatim}
shows that the variance of the resulting estimator is about $10^{-13}$. A simple simulation of a normal sample of size $m$ and the
resulting accounting of the portion of the sample above $4.5$ leads to a binomial estimator with a variance of $\mathbb{P}(Z>4.5)
\mathbb{P}(Z<4.5)/m$, which results in a lower bound
$$
m \ge \mathbb{P}(Z>4.5) \mathbb{P}(Z<4.5) / 1.5 10^{-13} \approx 0.75 10^{7}\,,
$$
i.e.~close to ten million simulations.
\end{enumerate}

\subsection{Exercise \ref{pb:fitz}}

For the three choices, the importance weights are easily computed:
\begin{verbatim}
x1=sample(c(-1,1),10^4,rep=T)*rexp(10^4)
w1=exp(-sqrt(abs(x1)))*sin(x1)^2*(x1>0)/.5*dexp(x1)
x2=rcauchy(10^4)*2
w2=exp(-sqrt(abs(x2)))*sin(x2)^2*(x2>0)/dcauchy(x2/2)
x3=rnorm(10^4)
w3=exp(-sqrt(abs(x3)))*sin(x3)^2*(x3>0)/dnorm(x3)
\end{verbatim}
They can be evaluated in many ways, from 
\begin{verbatim}
boxplot(as.data.frame(cbind(w1,w2,w3)))
\end{verbatim} 
to computing the effective sample size \verb=1/sum((w1/sum(w1))^2)= introduced in Example \ref{ex:probit}.
The preferable choice is then $g_1$. The estimated sizes are given by
\begin{verbatim}
> 4*10^6*var(x1*w1/sum(w1))/mean(x1*w1/sum(w1))^2
[1] 10332203
> 4*10^6*var(x2*w2/sum(w2))/mean(x2*w2/sum(w2))^2
[1] 43686697
> 4*10^6*var(x3*w3/sum(w3))/mean(x3*w3/sum(w3))^2
[1] 352952159
\end{verbatim}
again showing the appeal of using the double exponential proposal. (Note that efficiency could be
doubled by considering the absolute values of the simulations.)\\

\subsection{Exercise \ref{pb:top}}

\begin{enumerate}
\renewcommand{\theenumi}{\alph{enumi}}
\item With a positive density $g$ and the representation
$$ 
m(x) = \int_{\Theta}f(x|\theta)\dfrac{\pi(\theta)}{g(\theta)}g(\theta)\,\text{d}\theta\,,
$$
we can simulate $\theta_i$'s from $g$ to approximate $m(x)$ with
$$
\frac{1}{n}\sum_{i=1}^{n} \dfrac{f(x|\theta_{i})\pi(\theta_{i})}{g(\theta_{i})}\,. 
$$

\item When $ g(x) = \pi(\theta|x) = f(x|\theta)\pi(\theta)/K $, then
$$
K\frac{1}{n}\sum_{i=1}^{n} \dfrac{f(x|X_{i})\pi(X_{i})}{f(X_{i}|\theta)\pi(\theta)} = K 
$$
and the normalisation constant is the exact estimate. If the normalising constant is unknown, we
must use instead the self-normalising version \eqref{eq:Gby}.

\item Since
$$
\int_{\Theta} {\tau(\theta) \over f(x|\theta) \pi(\theta)} \pi(\theta|x) \text{d}\theta = 
\int_{\Theta} {\tau(\theta) \over f(x|\theta) \pi(\theta)} \dfrac{f(x|\theta) \pi(\theta)}{m(x)} 
\text{d}\theta = \dfrac{1}{m(x)}\,,
$$
we have an unbiased estimator of $1/m(x)$ based on simulations from the posterior,
$$
{1\over T} \sum_{t=1}^T {\tau(\theta_i^*) \over f(x|\theta_i^*) \pi(\theta_i^*)}
$$
and hence a converging (if biased) estimator of $m(x)$. This estimator of the marginal density can
then be seen as an harmonic mean estimator, but also as an importance sampling estimator \citep{robert:marin:2010}.
\end{enumerate}

\subsection{Exercise \ref{pb:margin}}

{\bf Warning: There is a typo in question b, which should read}
\begin{rema}
\noindent Let $X|Y=y \sim \CG (1,y)$ and $Y \sim \CE xp(1)$.
\end{rema}

\begin{enumerate}
\renewcommand{\theenumi}{\alph{enumi}}
\item If  $(X_i, Y_i) \sim f_{XY}(x,y)$, the Strong Law of Large Numbers tells us that
\begin{displaymath}
\lim_n {1\over n}
\sum_{i=1}^n \frac{f_{XY}(x^\ast, y_i) w(x_i)}{f_{XY}(x_i, y_i)}
= \int \int \frac{f_{XY}(x^\ast, y) w(x)}{f_{XY}(x, y)} f_{XY}(x,y) \text{d}x \text{d}y.
\end{displaymath}
Now cancel $f_{XY}(x,y)$ and use that fact that $\int w(x)dx=1$ to show
$$
\int \int \frac{f_{XY}(x^\ast, y) w(x)}{f_{XY}(x, y)} f_{XY}(x,y) \text{d}x \text{d}y=\int f_{XY}(x^\ast, y)  dy= f_X(x^\ast).
$$
\item  The exact marginal is 
$$
\int \left[y e^{-yx}\right] e^{-y} dy = \int y^{2-1} e^{-y(1+x)} dy = \frac{\gamma(2)}{(1+x)^2}.
$$
We tried the following \R version of Monte Carlo marginalization:
\begin{verbatim}
X=rep(0,nsim)
Y=rep(0,nsim)
for (i in 1:nsim){
   Y[i]=rexp(1)
   X[i]=rgamma(1,1,rate=Y[i])
   }

MCMarg=function(x,X,Y){
  return(mean((dgamma(x,1,rate=Y)/dgamma(X,1,
      rate=Y))*dgamma(X,7,rate=3)))
  }
True=function(x)(1+x)^(-2)
\end{verbatim}
which uses a $\mathcal{G}a(7,3)$ distribution to marginalize.  It works ok, as you
can check by looking at the plot 
\begin{verbatim}
> xplot=seq(0,5,.05);plot(xplot,MCMarg(xplot,X,Y)-True(xplot))
\end{verbatim}
\item Choosing $w(x) = f_{X}(x)$ leads to the estimator
\begin{align*}
\dfrac{1}{n} \sum_{i=1}^n \dfrac{f_{XY}(x^\ast, y_i) f_X(x_i)}{f_{XY}(x_i, y_i)}
&=
\dfrac{1}{n} \sum_{i=1}^n \dfrac{f_X(x^\ast)f_{Y|X}(y_i|x^\ast) f_X(x_i)}{f_X(x_i)f_{Y|X}(y_i|x_i)}
\\&= f_X(x^\ast)\, 
\dfrac{1}{n} \sum_{i=1}^n \dfrac{f_{Y|X}(y_i|x^\ast)}{f_{Y|X}(y_i|x_i)}
\end{align*}
which produces $f_X(x^\ast)$ modulo an estimate of $1$. If we decompose the variance of the estimator
in terms of 
$$
\text{var}\left\{\mathbb{E}\left[\left.\dfrac{f_{XY}(x^\ast, y_i) w(x_i)}{f_{XY}(x_i, y_i)}\right|x_i\right]\right\}+
\mathbb{E}\left\{\text{var}\left[\left.\dfrac{f_{XY}(x^\ast, y_i) w(x_i)}{f_{XY}(x_i, y_i)}\right|x_i\right]\right\}\,,
$$
the first term is 
\begin{align*}
\mathbb{E}\left[\left.\dfrac{f_{XY}(x^\ast, y_i) w(x_i)}{f_{XY}(x_i, y_i)}\right|x_i\right]&=
f_X(x^\ast)\mathbb{E}\left[\left.\dfrac{f_{Y|X}(y_i|x^\ast)}{f_{Y|X}(y_i|x_i)}\right|x_i\right]\,\dfrac{w(x_i)}{f_X(x_i)}\\
&= f_X(x^\ast)\dfrac{w(x_i)}{f_X(x_i)}
\end{align*}
which has zero variance if $w(x) = f_{X}(x)$. If we apply a variation calculus argument to the whole quantity, we
end up with
$$
w(x) \propto f_X(x) \bigg/ \int \dfrac{f^2_{Y|X}(y|x^\ast)}{f_{Y|X}(y|x)}\,\text{d}y
$$
minimizing the variance of the resulting estimator. So it is likely $f_X$ is {\em not} optimal...
\end{enumerate}

\chapter{Controling and Accelerating Convergence}

\subsection{Exercise \ref{pb:ratio_csts}}

\begin{enumerate}
\renewcommand{\theenumi}{\alph{enumi}}
\item Since
$$
\pi_1(\theta|x) = \tilde\pi_1(\theta)/c_1
                    \mbox{ and }\pi_2(\theta|x) =\tilde\pi_2(\theta)/c_2\,,
$$
where only $\tilde\pi_1$ and $\tilde\pi_2$ are known and where $c_1$ and $c_2$ correspond to
the marginal likelihoods, $m_1(x)$ and $m_2(x)$ (the dependence on $x$ is removed for simplification purposes),
we have that
$$
\varrho=\dfrac{m_1(x)}{m_2(x)}
=\dfrac{\int_{\Theta_1} \pi_1(\theta) f_1(x|\theta)\,\text{d}\theta}{\int_{\Theta_1} \pi_2(\theta) f_2(x|\theta)\,\text{d}\theta}
=\int_{\Theta_1} \dfrac{\pi_1(\theta) f_1(x|\theta)}{\tilde\pi_2(\theta)}\,\frac{\tilde\pi_2(\theta)}{m_2(x)}\text{d}\theta_1 
$$
and therefore $\tilde\pi_1(\theta)/\tilde\pi_2(\theta)$ is an unbiased estimator of $\varrho$ when $\theta\sim\pi_2(\theta|x)$.

\item Quite similarly,
$$
\dfrac{\int \tilde\pi_1(\theta) \alpha(\theta) \pi_2(\theta|x) \text{d}\theta }{
\int \tilde\pi_2(\theta) \alpha(\theta) \pi_1(\theta|x) \text{d}\theta} = 
\dfrac{\int \tilde\pi_1(\theta) \alpha(\theta) \tilde\pi_2(\theta)/c_2 \text{d}\theta }{
\int \tilde\pi_2(\theta) \alpha(\theta) \tilde\pi_1(\theta)/c_1 \text{d}\theta} = \frac{c_1}{c_2} = \varrho\,.
$$
\end{enumerate}

\subsection{Exercise \ref{exo:ESSin}}

We have
\begin{align*}
\text{ESS}_{n} &=1\bigg/\sum_{i=1}^{n}\underline{w}_{i}^{2}
=1\bigg/\sum_{i=1}^{n}\left(w_{i}\bigg/\sum_{j=1}^{n}w_{j}\right)^{2}\\
&=\dfrac{\left(\sum_{i=1}^{n}w_{i}\right)^{2}}{\sum_{i=1}^{n}w_{i}^{2}}
=\dfrac{\sum_{i=1}^{n}w_{i}^2+\sum_{i\neq j}w_{i}w_{j}}{\sum_{i=1}^{n}w_{i}^{2}}
\le n
\end{align*}
(This is also a consequence of Jensen's inequality when considering that the $\underline{w}_{i}$ sum up to one.)
Moreover, the last equality shows that
\[
ESS_{n}=1+\frac{\sum_{i\neq j}w_{i}w_{j}}{\sum_{i=1}^{n}w_{i}^{2}}\ge 1\,,
\]
with equality if and only if a single $\omega_i$ is different from zero.

\subsection{Exercise \ref{exo:simerin}}

{\bf Warning: There is a slight typo in the above in that $\bar {\mathbf X}_k$ should not be in bold. It should thus read}
\begin{rema}
\noindent Establish that
$$
\text{cov}(\bar {X}_k,\bar { X}_{k^\prime}) = {\sigma^2}\big/{\max\{k, k^\prime\}}.
$$
\end{rema}

Since the $X_{i}$'s are iid, for $k'<k$, we have
\begin{align*}
\text{cov}(\overline{X}_{k},\overline{X}_{k'}) 
 & = \text{cov}\left(\frac{1}{k}\sum_{i=1}^{k}X_{i},\frac{1}{k'}\sum_{i=1}^{k'}X_{i}\right)\\
 & = \text{cov}\left(\frac{1}{k}\sum_{i=1}^{k'}X_{i},\frac{1}{k'}\sum_{i=1}^{k'}X_{i}\right)\\
 & = \frac{1}{kk'}\text{cov}\left(\sum_{i=1}^{k'}X_{i},\sum_{i=1}^{k'}X_{i}\right)\\
 & = \frac{1}{kk'}k'\text{cov}\left(X_{i},X_{i}\right)\\
 & = \sigma^{2}/k\\
 & = \sigma^{2}/\max\{k,k'\}\,.
\end{align*}

\subsection{Exercise \ref{pb:t_RB}}

{\bf Warning: There is a missing variance term in this exercise, which should read}
\begin{rema}
\noindent Show that
\begin{eqnarray*}
\mathbb{E} \left[\exp-X^2|y\right] &=& \frac{1}{\sqrt{2 \pi \sigma^2/y}}
\int\exp\{-x^2\}\,\exp\{-(x-\mu)^2y/2\sigma^2\}\,\text{d}x \\
&=& \frac{1}{\sqrt{2 \sigma^2/y+1}} \exp \left\{ -\frac{\mu^2}{1+2\sigma^2/y}\right\}
\end{eqnarray*}
by completing the square in the exponent to evaluate the integral.
\end{rema}

We have
\begin{align*}
2x^2+(x-\mu)^2y/2\sigma^2 &= x^2(2+y\sigma^{-2}) -2x\mu y\sigma^{-2} + \mu^2 y\sigma^{-2}\\
&= (2+y\sigma^{-2})\left[x-\mu y\sigma^{-2}/(2+y\sigma^{-2})\right]^2+ \\
&\qquad\mu^2 \left[y\sigma^{-2}- y^2\sigma^{-4}/(2+y\sigma^{-2})\right]\\
&= (2+y\sigma^{-2})\left[x-\mu /(1+2\sigma^{2}/y)\right]^2+ \mu^2 /(1+2\sigma^{2}/y)
\end{align*}
and thus
\begin{align*}
\int \exp\{-x^2\}\,&\exp\{-(x-\mu)^2y/2\sigma^2\}\,\dfrac{\text{d}x}{\sqrt{2\pi\sigma^2/y}}\\
&=  \exp\left\{ -\frac{\mu^2}{1+2\sigma^2/y}\right\}\\
&\quad\times\int \exp\left\{-(2+y\sigma^{-2})\left[x-\mu /(1+2\sigma^{2}/y)\right]^2/2\right\}
\,\dfrac{\text{d}x}{\sqrt{2\pi\sigma^2/y}}\\
&= \exp \left\{ -\frac{\mu^2}{1+2\sigma^2/y}\right\}\,\dfrac{\sqrt{y\sigma^{-2}}}{\sqrt{2+y\sigma^{-2}}}\\
&= \exp \left\{ -\frac{\mu^2}{1+2\sigma^2/y}\right\}\,\dfrac{1}{\sqrt{1+2\sigma^{2}/y}}
\end{align*}

\subsection{Exercise \ref{exo:motown}}

Since $H(U)$ and $H(1_U)$ take opposite values when $H$ is monotone, i.e.~one is large when the other is small, those
two random variables are negatively correlated.

\subsection{Exercise \ref{pb:ratio_csts3} }

{\bf Warning: Another reference problem in this exercise: Exercise \ref{pb:ratio_csts2} should be Exercise \ref{pb:ratio_csts}.}

\begin{enumerate}
\renewcommand{\theenumi}{\alph{enumi}}
\item The ratio \eqref{eq:bridge} is a ratio of convergent estimators of the numerator and the denominator in question b of
Exercise \ref{pb:ratio_csts} when $\theta_{1i}\sim \pi_1(\theta|x)$ and $\theta_{2i} \sim \pi_2(\theta|x)$. (Note that the wording
of this question is vague in that it does not indicate the dependence on $x$.)
\item If we consider the special choice $\alpha(\theta) = 1 / \tilde\pi_1(\theta) \tilde\pi_2(\theta)$ in the representation of
question b of Exercise \ref{pb:ratio_csts}, we do obtain $\varrho = \BE^{\pi_2} [ \tilde\pi_2(\theta) ^{-1} ] / \BE^{\pi_1} 
[ \tilde\pi_1(\theta) ^{-1} ]$, assuming both expectations exist. Given that $(i=1,2)$
$$
\BE^{\pi_i} [ \tilde\pi_i(\theta) ^{-1} ] = \int_{\Theta} \dfrac{1}{\tilde\pi_i(\theta)}\,\dfrac{\tilde\pi_i(\theta)}{m_i(x)}\,
\text{d}\theta\,,
$$
this implies that the space $\Theta$ must have a finite measure. If $\text{d}\theta$ represents the dominating measure, $\Theta$
is necessarily compact.
\end{enumerate}
	
\subsection{Exercise \ref{pb:RBMix}}

Each of those \R programs compare the range of the Monte Carlo estimates with and without Rao--Blackwellization:
\begin{enumerate}
\renewcommand{\theenumi}{\alph{enumi}}
\item  For the negative binomial mean, $\BE_f(X)=a/b$ since $X\sim\mathcal{N}eg(a,b/(b+1))$.
\begin{verbatim}
y=matrix(rgamma(100*Nsim,a)/b,ncol=100)
x=matrix(rpois(100*Nsim,y),ncol=100)
matplot(apply(x,2,cumsum)/(1:Nsim),type="l",col="grey80",
lty=1,ylim=c(.4*a/b,2*a/b), xlab="",ylab="")
matplot(apply(y,2,cumsum)/(1:Nsim),type="l",col="grey40",
lty=1,add=T,xlab="",ylab="")
abline(h=a/b,col="gold",lty=2,lwd=2)
\end{verbatim}

\item For the generalized $t$ variable, $\BE_f(X)=BE_f[X|Y]=0$. So the improvement is obvious. To make a
more sensible comparison, we consider instead $\BE_f[X^2]=\BE[Y]=a/b$.
\begin{verbatim}
y=matrix(rgamma(100*Nsim,a)/b,ncol=100)
x=matrix(rnorm(100*Nsim,sd=sqrt(y)),ncol=100)
matplot(apply(x^2,2,cumsum)/(1:Nsim),type="l",col="grey80",
lty=1,ylim=(a/b)*c(.2,2), xlab="",ylab="")
matplot(apply(y,2,cumsum)/(1:Nsim),type="l",col="grey40",
lty=1,add=T,xlab="",ylab="")
abline(h=a/b,col="gold",lty=2,lwd=2)
\end{verbatim}

\item {\bf Warning: There is a typo in this question with a missing $n$ in the $\mathcal{B}in(y)$
distribution... It should be} 
\begin{rema}
c. $X|y \sim \mathcal{B}in(n,y)$, $Y \sim \mathcal{B}e(a,b)$ ($X$ is beta-binomial).
\end{rema}
In this case, $\BE_f[X]=n\BE_f[Y]=na/(a+b)$.
\begin{verbatim}
y=1/matrix(1+rgamma(100*Nsim,b)/rgamma(100*Nsim,a),ncol=100)
x=matrix(rbinom(100*Nsim,n,prob=y),ncol=100)
matplot(apply(x,2,cumsum)/(1:Nsim),type="l",col="grey80",lty=1,
ylim=(n*a/(a+b))*c(.2,2), xlab="",ylab="")
matplot(n*apply(y,2,cumsum)/(1:Nsim),type="l",col="grey40",lty=1,add=T,
xlab="",ylab="")
abline(h=n*a/(a+b),col="gold",lty=2,lwd=2)
\end{verbatim}

\end{enumerate}

\subsection{Exercise \ref{pb:bandaid}}

It should be clear from display (\ref{eq:sigmatri}) that we only need to delete the 
$n^2$ ($k^2$ in the current notation).  We replace it with $2 k^2$ and add the 
last row and column as in (\ref{eq:sigmatri}).

\subsection{Exercise \ref{pb:term_is}}

\begin{enumerate}
\renewcommand{\theenumi}{\alph{enumi}}
\item For the accept-reject algorithm,
\begin{eqnarray*}
&&\left(X_{1},\ldots,X_{m}\right)\sim f(x)\\
&&\left(U_{1},\ldots,U_{N}\right)\stackrel{\hbox{i.i.d.}}{\sim} \mathcal{U}_{[0,1]}\\
&&\left(Y_{1},\ldots,Y_{N}\right)\stackrel{\hbox{i.i.d.}}{\sim} g(y)\\
\end{eqnarray*}
and the acceptance weights are the $w_{j}=\frac{f\left(Y_{j}\right)}
{Mg\left(Y_{j}\right)}$. $N$ is the stopping time associated with these variables,
that is, $Y_{N}=X_{m}$. We have
\begin{eqnarray*}
\rho_{i}&=&P\left(U_{i}\leq w_{i}|N=n, Y_{1},\ldots,Y_{n}\right)\\
&=&\frac{P\left(U_{i}\leq w_{i},N=n,
Y_{1},\ldots,Y_{n}\right)}{P\left(N=n, Y_{1},\ldots,Y_{n}\right)}
\end{eqnarray*}
where the numerator is the probability that $Y_{N}$ is accepted as
$X_m$, $Y_{i}$ is accepted as one $X_j$ and there are $(m-2)$ $X_j$'s that
are chosen from the remaining $(n-2)$ $Y_\ell$'s. Since
\begin{equation*}
P\left(\hbox{$Y_{j}$ is accepted}\right)=P\left(U_{j}\leq w_{j}\right)=w_{j}\,,
\end{equation*}
the numerator is
\begin{equation*}
w_{i}\sum_{(i_{1},\ldots,i_{m-2})}\prod_{j=1}^{m-2}w_{i_{j}}\prod_{j=m-1}^{n-2}(1-w_{i_{j}})
\end{equation*}
where
\begin{enumerate}
\renewcommand{\theenumii}{\roman{enumii}}

\item $\prod_{j=1}^{m-2}w_{i_{j}}$ is the probability that among the
$N$ $Y_{j}$'s, in addition to both $Y_{N}$ and $Y_{i}$ being accepted,
there are $(m-2)$ other $Y_{j}$'s accepted as $X_\ell$'s;

\item $\prod_{j=m-1}^{n-2}(1-w_{i_{j}})$ is the probability
that there are $(n-m)$ rejected $Y_{j}$'s, given that
$Y_{i}$ and $Y_{N}$ are accepted;

\item the  sum is over all subsets of
$(1,\ldots,i-1,i+1,\ldots,n)$ since, except for $Y_{i}$ and $Y_{N}$,
other $(m-2)$ $Y_{j}$'s are chosen uniformly from $(n-2)$ $Y_{j}$'s.
\end{enumerate}

\par\noindent Similarly the denominator
$$
P\left(N=n,Y_{1},\ldots,Y_{n}\right)
=w_{i}\sum_{(i_{1},\ldots,i_{m-1})}\prod_{j=1}^{m-1}w_{i_{j}}\prod_{j=m}^{n-1}(1-w_{i_{j}})
$$
is the probability that $Y_{N}$ is accepted as $X_{m}$ and
$(m-1)$ other $X_{j}$'s are chosen from $(n-1)$ $Y_{\ell}$'s. Thus
\begin{eqnarray*}
\rho_{i}&=&P\left(U_{i}\leq w_{i}|N=n,Y_{1},\ldots,Y_{n}\right)\\
&=&w_{i}\frac{\sum_{(i_{1},\ldots,i_{m-2})}\prod_{j=1}^{m-2}w_{i_{j}}\prod_{j=m-1}^{n-2}(1-w_{i_{j}})}
{\sum_{(i_{1},\ldots,i_{m-1})}\prod_{j=1}^{m-1}w_{i_{j}}\prod_{j=m-1}^{n-1}(1-w_{i_{j}})}
\end{eqnarray*}

\item We have
\begin{eqnarray*}
\delta_{1}&=&\frac{1}{m}\sum_{i=1}^{m}h\left(X_{i}\right)=
\frac{1}{m}\sum_{j=1}^{N}h\left(Y_{j}\right)\mathbb{I}_{U_{j}\leq
w_{j}}\\
\delta_{2}&=&\frac{1}{m}\sum_{j=1}^{N}\mathbb{E}\left(\mathbb{I}_{U_{j}\leq
w_{j}}|N,Y_{1},\ldots,Y_{N}\right)h\left(Y_{j}\right)=
\frac{1}{m}\sum_{i=1}^{N}\rho_{i}h\left(Y_{i}\right)
\end{eqnarray*}

\par\noindent Since
$\mathbb{E}\left(\mathbb{E}\left(X|Y\right)\right)=\mathbb{E}\left(X\right)$,
\begin{eqnarray*}
\mathbb{E}\left(\delta_{2}\right)&=&\mathbb{E}\left(\frac{1}{m}\sum_{j=1}^{N}
\mathbb{E}\left(\mathbb{I}_{U_{j}\leq
w_{j}}|N,Y_{1},\ldots,Y_{N}\right)\right)\\
&=&\frac{1}{m}\sum_{j=1}^{N}\mathbb{E}\left(\mathbb{I}_{U_{j}\leq
w_{j}}\right)h\left(Y_{j}\right)\\
&=&\mathbb{E}\left(\frac{1}{m}\sum_{j=1}^{N}h\left(Y_{j}\right)\mathbb{I}_{U_{j}\leq
w_{j}}\right)=\mathbb{E}\left(\delta_{1}\right)
\end{eqnarray*}
Under quadratic loss, the risk of $\delta_{1}$ and $\delta_{2}$ are:
\begin{eqnarray*}
R\left(\delta_{1}\right)&=&\mathbb{E}\left(\delta_{1}-\mathbb{E}h\left(X\right)\right)^{2}\\
&=&\mathbb{E}\left(\delta_{1}^{2}\right)+\mathbb{E}\left(\mathbb{E}(h(X))\right)^{2}-
2\mathbb{E}\left(\delta_{1}\mathbb{E}(h(X))\right)\\
&=&\mathrm{var}\left(\delta_{1}\right)-\left(\mathbb{E}(\delta_{1})\right)^{2}+
\mathbb{E}\left(\mathbb{E}\left(h(X)\right)\right)^{2}-
2\mathbb{E}\left(\delta_{1}\mathbb{E}\left(h(X)\right)\right)
\end{eqnarray*}
and
\begin{eqnarray*}
R\left(\delta_{2}\right)&=&\mathbb{E}\left(\delta_{2}-\mathbb{E}h\left(X\right)\right)^{2}\\
&=&\mathbb{E}\left(\delta_{2}^{2}\right)+\mathbb{E}\left(\mathbb{E}(h(X))\right)^{2}-
2\mathbb{E}\left(\delta_{2}\mathbb{E}(h(X))\right)\\
&=&\mathrm{var}\left(\delta_{2}\right)-\left(\mathbb{E}(\delta_{2})\right)^{2}+
\mathbb{E}\left(\mathbb{E}\left(h(X)\right)\right)^{2}-
2\mathbb{E}\left(\delta_{2}\mathbb{E}\left(h(X)\right)\right)
\end{eqnarray*}
Since
$\mathbb{E}\left(\delta_{1}\right)=\mathbb{E}\left(\delta_{2}\right)$,
we only need to compare $\mathrm{var}\left(\delta_{1}\right)$ and
$\mathrm{var}\left(\delta_{2}\right)$. From the definition of
$\delta_{1}$ and $\delta_{2}$, we have
\begin{equation*}
\delta_{2}(X)=\mathbb{E}\left(\delta_{1}(X)|Y\right)
\end{equation*}
so
\begin{equation*}
\mathrm{var}\left(\mathbb{E}\left(\delta_{1}\right)\right)=\mathrm{var}\left(\delta_{2}\right)
\leq \mathrm{var}\left(\delta_{1}\right)\,.
\end{equation*}
\end{enumerate}

\subsection{Exercise \ref{pb:rob}}

\begin{enumerate}
\renewcommand{\theenumi}{\alph{enumi}}

\item Let us transform $\mathfrak{I}$ into $\mathfrak{I}=\int{\frac{h(y)f(y)}{m(y)}m(y)dy}$,
where $m$ is the marginal density of $Y_1$. We have
\begin{eqnarray*}
\mathfrak{I} &=& \sum_{n\in\N}{\left[P(N=n)\int{\frac{h(y)f(y)}{m(y|N=n)}}\right]}\\
&=& \mathbb{E}_N\left[\mathbb{E}\left[\frac{h(y)f(y)}{m(y)}|N\right]\right].
\end{eqnarray*}

\item As $\beta$ is constant, for every function $c$,
$$\mathfrak{I}=\beta\mathbb{E}[c(Y)]+\mathbb{E}\left[\frac{h(Y)f(Y)}{m(Y)}-\beta c(Y)\right].$$

\item The variance associated with an empirical mean of the
$$
\frac{h(Y_i)f(Y_i)}{m(Y_i)}-\beta c(Y_i)
$$
is
\begin{eqnarray*}
\mathrm{var}(\widehat{\mathfrak{I}}) &=& \beta^2\mathrm{var}(c(Y))+\mathrm{var}
\left(\frac{h(Y)f(Y)}{m(Y)}\right)-2\beta \mathrm{cov}\left[\frac{h(Y)
f(Y)}{m(Y)},c(Y)\right]\\
&=& \beta^2\mathrm{var}(c(Y))-2\beta
\mathrm{cov}[d(Y),c(Y)]+\mathrm{var}(d(Y)).
\end{eqnarray*}
Thus, the optimal choice of $\beta$ is such that
$$\frac{\partial \mathrm{var}(\widehat{\mathfrak{I}})}{\partial \beta}=0$$
and is given by
$$\beta^*=\frac{\mathrm{cov}[d(Y),c(Y)]}{\mathrm{var}(c(Y))}.$$

\item The first choice of $c$ is $c(y)={\mathbb{I}}_{\{y>y_0\}}$,
which is interesting when $p=P(Y>y_0)$ is known. In this case,
$$\beta^*=\frac{\int_{y>y_0}{hf}-\int_{y>y_0}{hf}\int_{y>y_0}{m}}{\int_{y>y_0}{m}-(\int_{y>y_0}{m})^2}=\frac{\int_{y>y_0}{hf}}{p}.$$
Thus, $\beta^*$ can be estimated using the Accept-reject sample. A
second choice of $c$ is $c(y)=y$, which leads to the two first
moments of $Y$. When those two moments $m_1$ and $m_2$ are known
or can be well approximated, the optimal choice of $\beta$ is
$$\beta^*=\frac{\int{yh(y)f(y)dy}-\mathfrak{I}m_1}{m_2}.$$
and can be estimated using the same sample or another instrumental
density namely when $\mathfrak{I}'=\int{yh(y)f(y)dy}$ is simple to
compute, compared to $\mathfrak{I}$.
\end{enumerate}

\chapter{Monte Carlo Optimization}

\subsection{Exercise \ref{exo:smplmix}}

\begin{figure}
\centerline{\includegraphics[width=\textwidth,height=5truecm]{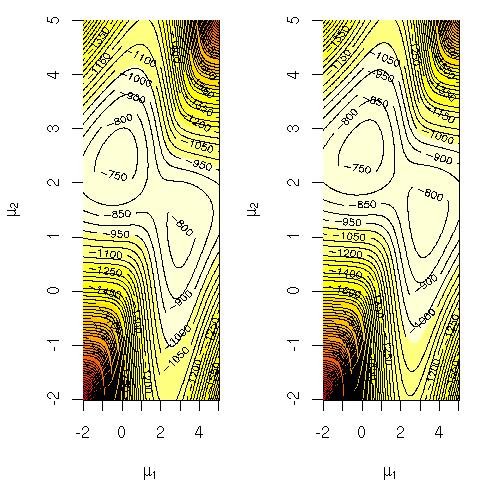}}
\caption{\label{fig:compamix}
Comparison of two log-likelihood surfaces for the mixture model \eqref{eq:maxmix}
when the data is simulated with a fixed $100/300$ ratio in both components {\em (left)}
and when the data is simulated with a binomial $\mathcal{B}(400,1/4)$ random number of points
on the first component.}
\end{figure}

This is straightforward in \R
\begin{verbatim}
par(mfrow=c(1,2),mar=c(4,4,1,1))
image(mu1,mu2,-lli,xlab=expression(mu[1]),ylab=expression(mu[2]))
contour(mu1,mu2,-lli,nle=100,add=T)
Nobs=400
da=rnorm(Nobs)+2.5*sample(0:1,Nobs,rep=T,prob=c(1,3))
for (i in 1:250)
for (j in 1:250)
  lli[i,j]=like(c(mu1[i],mu2[j]))
image(mu1,mu2,-lli,xlab=expression(mu[1]),ylab=expression(mu[2]))
contour(mu1,mu2,-lli,nle=100,add=T)
\end{verbatim}
Figure \ref{fig:compamix} shows that the log-likelihood surfaces are quite comparable, despite being
based on different samples. Therefore the impact of allocating $100$ and $300$ points to both components,
respectively, instead of the random $79$ and $321$ in the current realisation, is inconsequential. 

\subsection{Exercise \ref{ex:simpleMCO}}

{\bf Warning: as written, this problem has not simple solution! The constraint should be replaced with}
\begin{rema}
$$
x^2(1+\sin(y/3)\cos(8x))+y^2(2+\cos(5x)\cos(8y)) \le 1\,,
$$
\end{rema}

\begin{figure}
\centerline{\includegraphics[width=.75\textwidth]{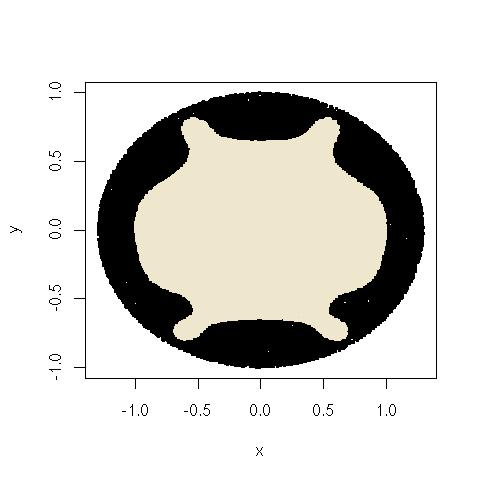}}
\caption{\label{fig:alien}
Simulation of a uniform distribution over a complex domain via uniform simulation over a
simpler encompassing domain for $10^5$ simulations and an acceptance rate of $0.73\% $.}
\end{figure}

We need to find a lower bound on the function of $(x,y)$. The coefficient of $y^2$ is obviously bounded
from below by $1$, while the coefficient of $x^2$ is positive. Since the function is bounded from below by $y^2$,
this means that $y^2<1$, hence that $\sin(y/3)>\sin(-1/3)>-.33$. Therefore, a lower bound on the function is $0.77x^2+y^2$. 
If we simulate uniformly over the ellipse $0.77x^2+y^2<1$, we can subsample the points that satisfy the constraint.
Simulating the uniform distribution on $0.77x^2+y^2<1$ is equivalent to simulate the uniform distribution over the unit
circle $z^2+y^2<1$ and resizing $z$ into $x=z/\sqrt{0.77}$.
\begin{verbatim}
theta=runif(10^5)*2*pi
rho=runif(10^5)
xunif=rho*cos(theta)/.77
yunif=rho*sin(theta)
plot(xunif,yunif,pch=19,cex=.4,xlab="x",ylab="y")
const=(xunif^2*(1+sin(yunif/3)*cos(xunif*8))+
 yunif^2*(2+cos(5*xunif)*cos(8*yunif))<1)
points(xunif[const],yunif[const],col="cornsilk2",pch=19,cex=.4)
\end{verbatim}
While the ellipse is larger than the region of interest, Figure \ref{fig:alien} shows that it is
reasonably efficient. The performances of the method are given by \verb+sum(const)/10^4+, which is
equal to $73\%$.

\subsection{Exercise \ref{exo:stogramix}}

Since the log-likelihood of the mixture model in Example \ref{ex:maxmix} has been defined by
\begin{verbatim}
#minus the log-likelihood function
like=function(mu){
  -sum(log((.25*dnorm(da-mu[1])+.75*dnorm(da-mu[2]))))
  }
\end{verbatim}
in the {\sf mcsm} package, we can reproduce the \R program of Example \ref{ex:find_max3} with the
function $h$ now defined as \verb+like+. The difference with the function $h$ of Example \ref{ex:find_max3} 
is that the mixture log-likelihood is more variable and thus the factors $\alpha_j$ and $\beta_j$ need to
be calibrated against divergent behaviours. The following figure shows the impact of the different choices
$(\alpha_j,\beta_j)=(.01/\log(j+1),1/\log(j+1)^{.5})$,
$(\alpha_j,\beta_j)=(.1/\log(j+1),1/\log(j+1)^{.5})$,
$(\alpha_j,\beta_j)=(.01/\log(j+1),1/\log(j+1)^{.1})$,
$(\alpha_j,\beta_j)=(.1/\log(j+1),1/\log(j+1)^{.1})$,
on the convergence of the gradient optimization. In particular, the second choice exhibits a particularly
striking behavior where the sequence of $(\mu_1,\mu_2)$ skirts the true mode of the likelihood in a circular
manner. (The stopping rule used in the \R program is \verb@(diff<10^(-5))@.)
\begin{figure}
\centerline{\includegraphics[width=.95\textwidth]{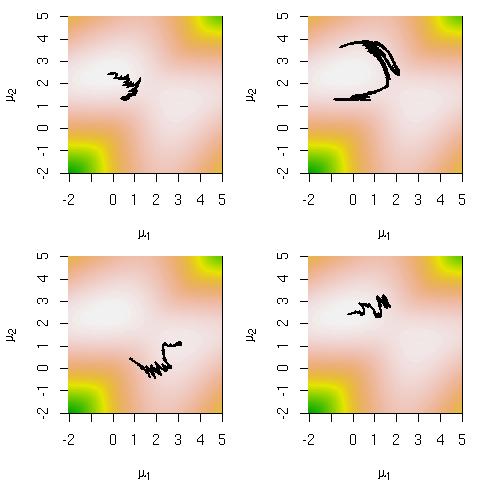}}
\caption{\label{fig:mixrad}
Four stochastic gradient paths for four different choices 
$(\alpha_j,\beta_j)=(.01/\log(j+1),1/\log(j+1)^{.5})$ (u.l.h.s.),
$(\alpha_j,\beta_j)=(.1/\log(j+1),1/\log(j+1)^{.5})$ (u.r.h.s.), 
$(\alpha_j,\beta_j)=(.01/\log(j+1),1/\log(j+1)^{.1})$ (l.l.h.s.),
$(\alpha_j,\beta_j)=(.1/\log(j+1),1/\log(j+1)^{.1})$ (l.r.h.s.).}
\end{figure}

\subsection{Exercise \ref{exo:freak}}

\begin{figure}
\centerline{\includegraphics[width=.95\textwidth]{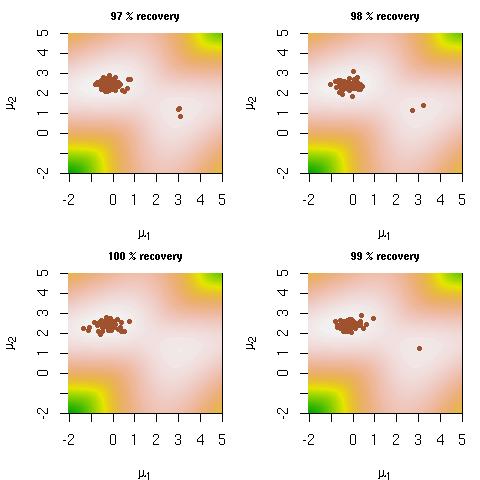}}
\caption{\label{fig:SAvabien}
Four simulated annealing outcomes corresponding to the temperature schedules
$T_t=1/1\log(1+t)$, 
$T_t=1/10\log(1+t)$, 
$T_t=1/10\sqrt{\log(1+t)}$, 
and $T_t=(.95)^{1+t}$, based on $100$ replications. (The percentage of recoveries of the main mode
is indicated in the title of each graph.)}
\end{figure}

The \R function \verb+SA+ provided in Example \ref{ex:mix_sa} can be used in the
following \R program to test whether or not the final value is closer to the main mode
or to the secondy mode:
\begin{verbatim}
modes=matrix(0,ncol=2,nrow=100)
prox=rep(0,100)
for (t in 1:100){
  res=SA(mean(da)+rnorm(2))
  modes[t,]=res$the[res$ite,]
  diff=modes[t,]-c(0,2.5)
  duff=modes[t,]-c(2.5,0)
  prox[t]=sum(t(diff)%*%diff<t(duff)%*%duff)
  }
\end{verbatim}

For each new temperature schedule, the function \verb0SA0 must be modified accordingly (for instance by
the on-line change \verb+SA=vi(SA)+). Figure \ref{fig:SAvabien} illustrates the output of an experiment
for four different schedules.

\subsection{Exercise 5.9}

In principle, $Q(\theta^\prime|\theta,\mathbf{x})$ should also involve the logarithms
of $1/4$ and $1/3$, raised to the powers $\sum Z_i$ and $\sum (1-Z_i)$, respectively.
But, due to the logarithmic transform, the expression does not involve the parameter
$\theta=(\mu_1,\mu_2)$ and can thus be removed from $Q(\theta^\prime|\theta,\mathbf{x})$
with no impact on the optimization problem.

\subsection{Exercise \ref{exo:tan1}}

{\bf Warning: there is a typo in Example \ref{ex:tan_wei}. The EM sequence should be
$$
\hat\theta_1 = \displaystyle{\left\{{\theta_0\,x_1\over 2+\theta_0}
+ x_4\right\}}\bigg/\displaystyle{\left\{{\theta_0\,x_1\over 2+\theta_0} +x_2+x_3+x_4\right\}} \;.
$$
instead of having $x_4$ in the denominator.}

Note first that some $1/4$ factors have been removed from every term as they were not
contributing to the likelihood maximisation. Given a starting point $\theta_0$, the 
EM sequence will always be the same.
\begin{verbatim}
x=c(58,12,9,13)
n=sum(x)
start=EM=cur=diff=.1
while (diff>.001){ #stopping rule

  EM=c(EM,((cur*x[1]/(2+cur))+x[4])/((cur*x[1]/(2+cur))+x[2]+x[3]+x[4]))
  diff=abs(cur-EM[length(EM)])
  cur=EM[length(EM)]
  }
\end{verbatim}
The Monte Carlo EM version creates a sequence based on a binomial simulation:
\begin{verbatim}
M=10^2
MCEM=matrix(start,ncol=length(EM),nrow=500)
for (i in 2:length(EM)){
  MCEM[,i]=1/(1+(x[2]+x[3])/(x[4]+rbinom(500,M*x[1],
  prob=1/(1+2/MCEM[,i-1]))/M))
  }
plot(EM,type="l",xlab="iterations",ylab="MCEM sequences")
upp=apply(MCEM,2,max);dow=apply(MCEM,2,min)
polygon(c(1:length(EM),length(EM):1),c(upp,rev(dow)),col="grey78")
lines(EM,col="gold",lty=2,lwd=2)
}
\end{verbatim}
and the associated graph shows a range of values that contains the true EM sequence. Increasing
\verb=M= in the above \R program obviously reduces the range.

\subsection{Exercise \ref{exo:maxmim}}

\begin{figure}
\centerline{\includegraphics[width=.8\textwidth]{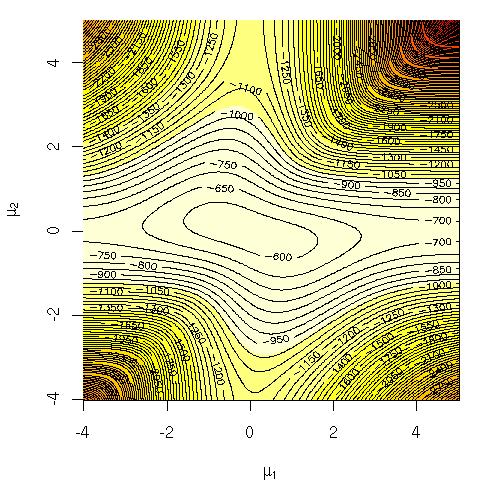}}
\caption{\label{fig:fakmix}
Log-likelihood surface of a mixture model applied to a five component mixture
sample of size $400$.}
\end{figure}

The \R function for plotting the (log-)likelihood surface associated
with \eqref{eq:maxmix} was provided in Example \ref{ex:maxmix}. 
We thus simply need to apply this function to the new sample,
resulting in an output like Figure \ref{fig:fakmix}, with a single mode
instead of the usual two modes. 

\subsection{Exercise \ref{pb:gyr_tre}}

{\bf Warning: there is a typo in question a where the formula should involve capital
$Z_i$'s, namely}
\begin{rema}
$$
P(Z_i=1) = 1 - P(Z_i=2) = { p \lambda \exp (-\lambda x_i) \over
p \lambda \exp (-\lambda x_i) +(1-p) \mu \exp (-\mu x_i)}.
$$
\end{rema}

\begin{enumerate}
\renewcommand{\theenumi}{\alph{enumi}}
\item The likelihood is
$$L(\theta|{\bf x})=\prod_{i=1}^{12}{[p\lambda e^{-\lambda x_i}+(1-p)\mu e^{-\mu x_i}]},$$
and the complete-data likelihood is
$$
L^c(\theta|{\bf x},{\bf z})=\prod_{i=1}^{12}{[p\lambda e^{-\lambda x_i}\mathbb{I}_{(z_i=1)}+(1-p)\mu e^{-\mu x_i}\mathbb{I}_{(z_i=2)}]},
$$
where $\theta=(p,\lambda,\mu)$ denotes the parameter, using the same arguments as in Exercise 
\ref{pb:7.4.5.1}.

\item The EM algorithm relies on the optimization of the expected log-likelihood
\begin{align*}
Q(\theta|\hat\theta_{(j)},{\bf x})&=\sum_{i=1}^{12} \left[\log{(p\lambda e^{-\lambda x_i})}
P_{\hat\theta_{(j)}}(Z_i=1|x_i)\right.\\
&\left.\quad +\log{((1-p)\mu e^{-\mu x_i})}P_{\hat\theta_{(j)}}(Z_i=2|x_i)\right].
\end{align*}
The arguments of the maximization problem are
$$
\left\{%
\begin{array}{lll}
\hat p_{(j+1)}=\hat P/12\\
\hat\lambda_{(j+1)}=\hat S_1/\hat P\\
\hat\mu_{(j+1)}=\hat S_2/\hat P,
\end{array}%
\right.
$$
where
$$
\left\{%
\begin{array}{lll}
\hat P=\sum_{i=1}^{12}{P_{\hat\theta_{(j)}}(Z_i=1|x_i)}\\\\
\hat S_1=\sum_{i=1}^{12}{x_iP_{\hat\theta_{(j)}}(Z_i=1|x_i)}\\\\
\hat S_2=\sum_{i=1}^{12}{x_iP_{\hat\theta_{(j)}}(Z_i=2|x_i)}\\
\end{array}%
\right.
$$
with
$$
P_{\hat\theta_{(j)}}(Z_i=1|x_i)
=\frac{\hat p_{(j)}\hat\lambda_{(j)}e^{-\hat\lambda_{(j)}x_i}}{\hat
p_{(j)}\hat\lambda_{(j)}e^{-\hat\lambda_{(j)}x_i}+(1-\hat
p_{(j)})\hat\mu_{(j)}e^{-\hat\mu_{(j)}x_i}}\,.
$$
An \R implementation of the algorithm is then
\begin{verbatim}
x=c(0.12,0.17,0.32,0.56,0.98,1.03,1.10,1.18,1.23,1.67,1.68,2.33)
EM=cur=c(.5,jitter(mean(x),10),jitter(mean(x),10))
diff=1
while (diff*10^5>1){
  
  probs=1/(1+(1-cur[1])*dexp(x,cur[3])/(cur[1]*dexp(x,cur[2])))
  phat=sum(probs);S1=sum(x*probs);S2=sum(x*(1-probs))
  EM=rbind(EM,c(phat/12,S1/phat,S2/phat))
  diff=sum(abs(cur-EM[dim(EM)[1],]))
  cur=EM[dim(EM)[1],]
  }
\end{verbatim}
and it always produces a single component mixture.
\end{enumerate}

\subsection{Exercise \ref{pb:EMCensored}}

{\bf Warning: Given the notations of Example \ref{ex:EMCensored2}, the function
$\phi$ in question b should be written $\varphi$...}
\begin{enumerate}
\renewcommand{\theenumi}{\alph{enumi}}
\item The question is a bit vague in that the density of the missing data $(Z_{n-m+1},\ldots,Z_n)$ is a normal
${\cal N}(\theta, 1)$ density if we do not condition on $\by$. Conditional upon $\by$, the missing observations 
$Z_i$ are truncated in $a$, i.e.~we know that they are larger than $a$. The conditional distribution of the $Z_i$'s
is therefore a normal ${\cal N}(\theta, 1)$ distribution truncated in $a$, with density
$$
f(z|\theta,y) = \dfrac{\exp\{-(z_i-\theta)^2/2\}}{\sqrt{2\pi}\,P_\theta(Y>a)}\,\mathbb{I}{z\ge a}\,.
= \dfrac{\varphi(z-\theta)}{1-\Phi(a-\theta)}\,\mathbb{I}{z\ge a}\,.
$$
where $\varphi$ and $\Phi$ are the normal pdf and cdf, respectively.
\item We have
\begin{align*}
\BE_{\theta}[Z_i|Y_i] &= \int_a^\infty z\,\dfrac{\varphi(z-\theta)}{1-\Phi(a-\theta)}\,\text{d}z\\
&= \theta + \int_a^\infty (z-\theta)\,\dfrac{\varphi(z-\theta)}{1-\Phi(a-\theta)}\,\text{d}z\\
&= \theta + \int_{a-\theta}^\infty y\,\dfrac{\varphi(y)}{1-\Phi(a-\theta)}\,\text{d}y\\
&= \theta + \left[-\varphi(x)\right]_{a-\theta}^\infty\\
&= \theta + \frac{\varphi(a-\theta)}{1-\Phi(a-\theta)},
\end{align*}
since $\varphi^\prime(x)=-x\varphi(x)$.
\end{enumerate}

\subsection{Exercise \ref{pb:uniroot}}

Running \verb+uniroot+ on both intervals
\begin{verbatim}
> h=function(x){(x-3)*(x+6)*(1+sin(60*x))}
> uniroot(h,int=c(-2,10))
$root
[1] 2.999996
$f.root
[1] -6.853102e-06
> uniroot(h,int=c(-8,1))
$root
[1] -5.999977
$f.root
[1] -8.463209e-06
\end{verbatim}
misses all solutions to $1+\sin(60x)=0$

\subsection{Exercise \ref{pb:used_up1}}

{\bf Warning: this Exercise duplicates Exercise \ref{exo:tan1} and should not
have been included in the book!}\\

\chapter{Metropolis-Hastings Algorithms}

\subsection{Exercise \ref{exo:AR}}

A simple \R program to simulate this chain is
\begin{verbatim}
# (C.) Jiazi Tang, 2009
x=1:10^4
x[1]=rnorm(1)
r=0.9
for (i in 2:10^4){
  x[i]=r*x[i-1]+rnorm(1) }
hist(x,freq=F,col="wheat2",main="")
curve(dnorm(x,sd=1/sqrt(1-r^2)),add=T,col="tomato"
\end{verbatim}

\subsection{Exercise \ref{exo:rho}}

When $q(y|x)=g(y)$, we have 
\begin{align*}
\rho(x,y) &= \min\left(\frac{f(y)}{f(x)} \frac{q(x|y)}{q(y|x)},1\right)\\
&= \min\left(\frac{f(y)}{f(x)} \frac{g(x)}{g(y)},1\right)\\
&= \min\left(\frac{f(y)}{f(x)} \frac{g(x)}{g(y)},1\right)\,.
\end{align*}
Since the acceptance probability satisfies
$$
\frac{f(y)}{f(x)} \frac{g(x)}{g(y)} \ge \frac{f(y)/g(y)}{\max f(x)/g(x)}
$$
it is larger for Metropolis--Hastings than for accept-reject.

\subsection{Exercise \ref{exo:mocho}}

\begin{enumerate}
\renewcommand{\theenumi}{\alph{enumi}}
\item The first property follows
from a standard property of the normal distribution, namely that the linear transform of a normal
is again normal. The second one is a consequence of the decomposition $y = X\beta + \epsilon$, when
$\epsilon\sim\mathcal{N}_n(0,\sigma^2 I_n)$ is independent from $X\beta$.  
\item This derivation is detailed in Marin and Robert (2007, Chapter 3, Exercise 3.9).

Since
$$
\by|\sigma^2,X\sim\mathcal{N}_n(X\tilde\beta,\sigma^2(I_n+n X(X^\text{T} X)^{-1}X^\text{T} ))\,,
$$
integrating in $\sigma^2$ with $\pi(\sigma^2)=1/\sigma^2$ yields
\begin{eqnarray*}
f(\by|X) & = & (n+1)^{-(k+1)/2}\pi^{-n/2}\Gamma(n/2)\left[\by^\text{T} \by
    -\frac{n}{n+1}\by^\text{T} X(X^\text{T} X)^{-1}X^\text{T} \by\right. \\
       &&\qquad  -\left.\frac{1}{n+1}\tilde\beta^\text{T} X^\text{T} X\tilde\beta\right]^{-n/2}.
\end{eqnarray*}
Using the \R function \verb+dmt(mnormt)+, we obtain the marginal density for the swiss dataset:
\begin{verbatim}
> y=log(as.vector(swiss[,1]))
> X=as.matrix(swiss[,2:6])
> library(mnormt)
> dmt(y,S=diag(length(y))+X%*%solve(t(X)%*%X)%*%t(X),d=length(y)-1)
[1] 2.096078e-63
\end{verbatim}
with the prior value $\tilde\beta=0$.
\end{enumerate}

\subsection{Exercise \ref{pb:beta}} 

\begin{enumerate}
\renewcommand{\theenumi}{\alph{enumi}}
\item We generate an \HM sample from the ${\cal B}e(2.7,6.3)$ density using uniform simulations:
\begin{verbatim}
# (C.) Thomas Bredillet, 2009
Nsim=10^4
a=2.7;b=6.3
X=runif(Nsim)
last=X[1]
for (i in 1:Nsim) {
        cand=rbeta(1,1,1)
        alpha=(dbeta(cand,a,b)/dbeta(last,a,b))/
	      (dbeta(cand,1,1)/dbeta(last,1,1))
        if (runif(1)<alpha) 
           last=cand
        X[i]=last
	}
hist(X,pro=TRUE,col="wheat2",xlab="",ylab="",main="Beta(2.7,3) simulation")
curve(dbeta(x,a,b),add=T,lwd=2,col="sienna2")
\end{verbatim}
The acceptance rate is estimated by
\begin{verbatim}
> length(unique(X))/5000
[1] 0.458
\end{verbatim}
If instead we use a ${\cal B}e(20,60)$ proposal, the modified lines in the \R program are
\begin{verbatim}
cand=rbeta(20,60,1)
alpha=(dbeta(cand,a,b)/dbeta(last,a,b))/
      (dbeta(cand,20,60)/dbeta(last,20,60))
\end{verbatim}
and the acceptance rate drops to zero! 

\item In the case of a truncated beta, the following \R program
\begin{verbatim}
Nsim=5000
a=2.7;b=6.3;c=0.25;d=0.75
X=rep(runif(1),Nsim)
test2=function(){
  last=X[1]
  for (i in 1:Nsim){
        cand=rbeta(1,2,6)
        alpha=(dbeta(cand,a,b)/dbeta(last,a,b))/
	      (dbeta(cand,2,6)/dbeta(last,2,6))
        if ((runif(1)<alpha)&&(cand<d)&&(c<cand)) 
           last=cand 
        X[i]=last}
  }
test1=function(){
  last=X[1]
  for (i in 1:Nsim){
        cand=runif(1,c,d)
        alpha=(dbeta(cand,a,b)/dbeta(last,a,b))
        if ((runif(1)<alpha)&&(cand<d)&&(c<cand))
           last=cand
        X[i]=last
        }
}
system.time(test1());system.time(test2())
\end{verbatim}
shows very similar running times but more efficiency for the beta proposal, since the
acceptance rates are approximated by $0.51$ and $0.72$ for \verb+test1+ and \verb+test2+,
respectively. When changing to $c=0.25$, $d=0.75$, \verb+test1+ is more efficient than \verb=test2=,
with acceptances rates of approximately $0.58$ and $0.41$, respectively.
\end{enumerate}

\subsection{Exercise \ref{pb:met_compare}}

\begin{enumerate}
\renewcommand{\theenumi}{\alph{enumi}}
\item  The \AR~algorithm with a Gamma $\CG(4,7)$ candidate can be implemented as follows
\begin{verbatim}
# (C.) Jiazi Tang, 2009
g47=rgamma(5000,4,7)
u=runif(5000,max=dgamma(g47,4,7))
x=g47[u<dgamma(g47,4.3,6.2)]
par(mfrow=c(1,3),mar=c(4,4,1,1))
hist(x,freq=FALSE,xlab="",ylab="",col="wheat2",
main="Accept-Reject with Ga(4.7) proposal")
curve(dgamma(x,4.3,6.2),lwd=2,col="sienna",add=T)
\end{verbatim}
The efficiency of the simulation is given by
\begin{verbatim}
> length(x)/5000
[1] 0.8374
\end{verbatim}

\item  The \HM~algorithm with a Gamma $\CG(4,7)$ candidate can be implemented as follows
\begin{verbatim}
# (C.) Jiazi Tang, 2009
X=rep(0,5000)
X[1]=rgamma(1,4.3,6.2)
for (t in 2:5000){
    rho=(dgamma(X[t-1],4,7)*dgamma(g47[t],4.3,6.2))/
        (dgamma(g47[t],4,7)*dgamma(X[t-1],4.3,6.2))
    X[t]=X[t-1]+(g47[t]-X[t-1])*(runif(1)<rho)
    }
hist(X,freq=FALSE,xlab="",ylab="",col="wheat2",
main="Metropolis-Hastings with Ga(4,7) proposal")
curve(dgamma(x,4.3,6.2),lwd=2,col="sienna",add=T)
\end{verbatim}
Its efficiency is
\begin{verbatim}
> length(unique(X))/5000
[1] 0.79
\end{verbatim}

\item The \MH~algorithm with a Gamma $\CG(5,6)$ candidate can be implemented as follows
\begin{verbatim}
# (C.) Jiazi Tang, 2009
g56=rgamma(5000,5,6)
X[1]=rgamma(1,4.3,6.2)
for (t in 2:5000){
   rho=(dgamma(X[t-1],5,6)*dgamma(g56[t],4.3,6.2))/
       (dgamma(g56[t],5,6)*dgamma(X[t-1],4.3,6.2))
   X[t]=X[t-1]+(g56[t]-X[t-1])*(runif(1)<rho)
   }
hist(X,freq=FALSE,xlab="",ylab="",col="wheat2",
main="Metropolis-Hastings with Ga(5,6) proposal")
curve(dgamma(x,4.3,6.2),lwd=2,col="sienna",add=T)
\end{verbatim}
Its efficiency is
\begin{verbatim}
> length(unique(X))/5000
[1] 0.7678
\end{verbatim}
which is therefore quite similar to the previous proposal.
\end{enumerate} 

\subsection{Exercise \ref{exo:brakin}}  

\begin{enumerate}
\renewcommand{\theenumi}{\arabic{enumi}.}
\item  Using the candidate given in Example \ref{ex:braking} mean using the \verb+Braking+ \R program of
our package \verb+mcsm+. In the earlier version, there is a missing link in the \R function which must
then be corrected by changing
\begin{verbatim}
data=read.table("BrakingData.txt",sep = "",header=T)
x=data[,1]
y=data[,2]
\end{verbatim}
into
\begin{verbatim}
x=cars[,1]
y=cars[,2]
\end{verbatim}
In addition, since the original \verb$Braking$ function does not return the simulated chains, a final line
\begin{verbatim}
list(a=b1hat,b=b2hat,c=b3hat,sig=s2hat)
\end{verbatim}
must be added into the function.
\item If we save the chains as \verb+mcmc=Braking()+ (note that we use $10^3$ simulations instead of $500$), 
the graphs assessing convergence can be plotted by
\begin{verbatim}
par(mfrow=c(3,3),mar=c(4,4,2,1))
plot(mcmc$a,type="l",xlab="",ylab="a");acf(mcmc$a) 
hist(mcmc$a,prob=T,main="",yla="",xla="a",col="wheat2")
plot(mcmc$b,type="l",xlab="",ylab="b");acf(mcmc$b) 
hist(mcmc$b,prob=T,main="",yla="",xla="b",col="wheat2")
plot(mcmc$c,type="l",xlab="",ylab="c");acf(mcmc$c) 
hist(mcmc$c,prob=T,main="",yla="",xla="c",col="wheat2")
\end{verbatim}
Autocorrelation graphs provided by \verb+acf+ show a strong correlation across iterations, while the raw plot
of the sequences show poor acceptance rates. The histograms are clearly unstable as well. This $10^3$ iterations
do not appear to be sufficient in this case.
\item Using 
\begin{verbatim}
> quantile(mcmc$a,c(.025,.975))
     2.5%     97.5%
-6.462483 12.511916
\end{verbatim}
and the same for $b$ and $c$ provides converging confidence intervals on the three parameters.
\end{enumerate}

\subsection{Exercise \ref{ex:challenger2}}

{\bf Warning: There is a typo in question b in that the candidate must also be a double-exponential for $\alpha$, since
there is no reason for $\alpha$ to be positive...}

\begin{figure}
\centerline{\includegraphics[width=\textwidth]{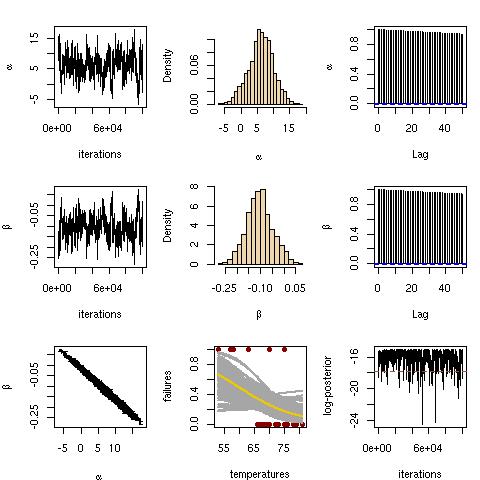}}
\caption{\label{fig:mhuttle}
Graphical checks of the convergence of the Metropolis--Hastings algorithm associated with
the {\sf challenger} dataset and a logistic regression model.}
\end{figure}

\begin{enumerate}
\renewcommand{\theenumi}{\arabic{enumi}}
\item The dataset {\tt challenger} is provided with the \verb+mcsm+ package, thus available as
\begin{verbatim}
> library(mcsm)
> data(challenger)
\end{verbatim}
Running a regular logistic regression is a simple call to \verb+glm+:
\begin{verbatim}
> temper=challenger[,2]
> failur=challenger[,1]
> summary(glm(failur~temper, family = binomial))

Deviance Residuals:
    Min       1Q   Median       3Q      Max
-1.0611  -0.7613  -0.3783   0.4524   2.2175

Coefficients:
            Estimate Std. Error z value Pr(>|z|)
(Intercept)  15.0429     7.3786   2.039   0.0415 *
temper       -0.2322     0.1082  -2.145   0.0320 *
---
Signif. codes:  0 "***" .001 "**" .01 "**" .05 "." .1 "" 1

(Dispersion parameter for binomial family taken to be 1)

    Null deviance: 28.267  on 22  degrees of freedom
Residual deviance: 20.315  on 21  degrees of freedom
AIC: 24.315
\end{verbatim}
The MLE's and the associated covariance matrix are given by
\begin{verbatim}
> challe=summary(glm(failur~temper, family = binomial))
> beta=as.vector(challe$coef[,1])
> challe$cov.unscaled
            (Intercept)      temper
(Intercept)  54.4441826 -0.79638547
temper       -0.7963855  0.01171512
\end{verbatim}
The result of this estimation can be checked by
\begin{verbatim}
plot(temper,failur,pch=19,col="red4",
xlab="temperatures",ylab="failures")
curve(1/(1+exp(-beta[1]-beta[2]*x)),add=TRUE,col="gold2",lwd=2)
\end{verbatim}
and the curve shows a very clear impact of the temperature.

\item The Metropolis--Hastings resolution is based on the \verb+challenge(mcsm)+ function, using the same
prior on the coefficients, $\alpha\sim\mathcal{N}(0,25)$, $\beta\sim\mathcal{N}(0,25/s^2_x)$, where $s^2_x$
is the empirical variance of the temperatures.
\begin{verbatim}
Nsim=10^4
x=temper
y=failur
sigmaa=5
sigmab=5/sd(x)

lpost=function(a,b){
  sum(y*(a+b*x)-log(1+exp(a+b*x)))+
  dnorm(a,sd=sigmaa,log=TRUE)+dnorm(b,sd=sigmab,log=TRUE)
   }

a=b=rep(0,Nsim)
a[1]=beta[1]
b[1]=beta[2]
#scale for the proposals
scala=sqrt(challe$cov.un[1,1])
scalb=sqrt(challe$cov.un[2,2])

for (t in 2:Nsim){
  propa=a[t-1]+sample(c(-1,1),1)*rexp(1)*scala
  if (log(runif(1))<lpost(propa,b[t-1])-
    lpost(a[t-1],b[t-1])) a[t]=propa
  else a[t]=a[t-1]
  propb=b[t-1]+sample(c(-1,1),1)*rexp(1)*scalb
  if (log(runif(1))<lpost(a[t],propb)-
    lpost(a[t],b[t-1])) b[t]=propb
  else b[t]=b[t-1]
  }
\end{verbatim}
The acceptance rate is low
\begin{verbatim}
> length(unique(a))/Nsim
[1] 0.1031
> length(unique(b))/Nsim
[1] 0.1006
\end{verbatim}
but still acceptable.
\item Exploring the output can be done via graphs as follows
\begin{verbatim}
par(mfrow=c(3,3),mar=c(4,4,2,1))
plot(a,type="l",xlab="iterations",ylab=expression(alpha))
hist(a,prob=TRUE,col="wheat2",xlab=expression(alpha),main="")
acf(a,ylab=expression(alpha))
plot(b,type="l",xlab="iterations",ylab=expression(beta))
hist(b,prob=TRUE,col="wheat2",xlab=expression(beta),main="")
acf(b,ylab=expression(beta))
plot(a,b,type="l",xlab=expression(alpha),ylab=expression(beta))
plot(temper,failur,pch=19,col="red4",
   xlab="temperatures",ylab="failures")
for (t in seq(100,Nsim,le=100)) curve(1/(1+exp(-a[t]-b[t]*x)),
  add=TRUE,col="grey65",lwd=2)
curve(1/(1+exp(-mean(a)-mean(b)*x)),add=TRUE,col="gold2",lwd=2.5)
postal=rep(0,1000);i=1
for (t in seq(100,Nsim,le=1000)){ postal[i]=lpost(a[t],b[t]);i=i+1}
plot(seq(100,Nsim,le=1000),postal,type="l",
  xlab="iterations",ylab="log-posterior")
abline(h=lpost(a[1],b[1]),col="sienna",lty=2)
\end{verbatim}
which shows a slow convergence of the algorithm (see the \verb+acf+ graphs on Figure \ref{fig:mhuttle}!)
\item The predictions of failure are given by
\begin{verbatim}
> mean(1/(1+exp(-a-b*50)))
[1] 0.6898612
> mean(1/(1+exp(-a-b*60)))
[1] 0.4892585
> mean(1/(1+exp(-a-b*70)))
[1] 0.265691
\end{verbatim}
\end{enumerate}

\subsection{Exercise \ref{pb:Norm-DE}}  

{\bf Warning: There is a typo in question c, which should involve $\mathcal{N}(0,\omega)$
candidates instead of $\mathcal{L}(0,\omega)$...}\\

\begin{enumerate}
\renewcommand{\theenumi}{\alph{enumi}}
\item An \R program to produce the three evaluations is
\begin{verbatim}
# (C.) Thomas Bredillet, 2009
Nsim=5000
A=B=runif(Nsim)
alpha=1;alpha2=3
last=A[1] 
a=0;b=1
cand=ifelse(runif(Nsim)>0.5,1,-1) * rexp(Nsim)/alpha
for (i in 1:Nsim){
        rate=(dnorm(cand[i],a,b^2)/dnorm(last,a,b^2))/
       (exp(-alpha*abs(cand[i]))/exp(-alpha*abs(last)))
        if (runif(1)<rate) last=cand[i]
        A[i]=last
        }
cand=ifelse(runif(Nsim)>0.5,1,-1) * rexp(Nsim)/alpha2
for (i in 1:Nsim) {
       rate=(dnorm(cand[i],a,b^2)/dnorm(last,a,b^2))/
       (exp(-alpha2*abs(cand[i]))/exp(-alpha2*abs(last)))
       if (runif(1)<rate) last=cand[i]
       B[i]=last
       }
par (mfrow=c(1,3),mar=c(4,4,2,1))
est1=cumsum(A)/(1:Nsim)
est2=cumsum(B)/(1:Nsim)
plot(est1,type="l",xlab="iterations",ylab="",lwd=2)
lines(est2,lwd="2",col="gold2")
acf(A)
acf(B)
\end{verbatim}

\item The acceptance rate is given by \verb+length(unique(B))/Nsim+, equal to
$0.49$ in the current simulation. A plot of the acceptance rates can be done
via the \R program
\begin{verbatim}
alf=seq(1,10,le=50)
cand0=ifelse(runif(Nsim)>0.5,1,-1) * rexp(Nsim)
acce=rep(0,50)
for (j in 1:50){
  cand=cand0/alf[j]
  last=A[1]
  for (i in 2:Nsim){
    rate=(dnorm(cand[i],a,b^2)/dnorm(last,a,b^2))/
    (exp(-alf[j]*abs(cand[i]))/exp(-alf[j]*abs(last)))
    if (runif(1)<rate) last=cand[i]
    A[i]=last
    }
  acce[j]=length(unique(A))/Nsim
  }
par(mfrow=c(1,3),mar=c(4,4,2,1))
plot(alf,acce,xlab="",ylab="",type="l",main="Laplace iid")
\end{verbatim}
The highest acceptance rate is obtained for the smallest value of $\alpha$.
   
\item The equivalent of the above \R program is
\begin{verbatim}
ome=sqrt(seq(.01,10,le=50))
cand0=rnorm(Nsim)
acce=rep(0,50)
for (j in 1:50){
  cand=cand0*ome[j]
  last=A[1]
  for (i in 2:Nsim){
    rate=(dnorm(cand[i],a,b^2)/dnorm(last,a,b^2))/
    (dnorm(cand[i],sd=ome[j])/dnorm(last,sd=ome[j]))
    if (runif(1)<rate) last=cand[i]
    A[i]=last
    }
  acce[j]=length(unique(A))/Nsim
  }
plot(ome^2,acce,xlab="",ylab="",type="l",main="Normal iid")
\end{verbatim}
The highest acceptance rate is (unsurprisingly) obtained for $\omega$ close to $1$.

\item The equivalent of the above \R program is
\begin{verbatim}
alf=seq(.1,10,le=50)
cand0=ifelse(runif(Nsim)>0.5,1,-1) * rexp(Nsim)
acce=rep(0,50)
for (j in 1:50){
  eps=cand0/alf[j]
  last=A[1]
  for (i in 2:Nsim){
    cand[i]=last+eps[i]
    rate=dnorm(cand[i],a,b^2)/dnorm(last,a,b^2)
    if (runif(1)<rate) last=cand[i]
    A[i]=last
    }
  acce[j]=length(unique(A))/Nsim
  }
plot(alf,acce,xlab="",ylab="",type="l",main="Laplace random walk")
\end{verbatim}
Unsurprisingly, as $\alpha$ increases, so does the acceptance rate. However, given that
this is a random walk proposal, higher acceptance rates do not mean better performances
(see Section 6.5).
\end{enumerate}

\chapter{Gibbs Samplers}

\newcommand{\EG}{Gibbs sampling$\;$}
\newcommand{\GR}{Gibbs sampler$\;$}
\subsection{Exercise \ref{exo:margikov}}

The density $g_{t}$ of $(X_{t},Y_{t})$ in Algorithm \ref{al:TSGibbs} is decomposed as
\begin{align*}
g_{t}(X_{t},Y_{t}|X_{t-1},&\dots X_{0},Y_{t-1},\dots Y_{0}) 
= g_{t,X|Y}(X_{t}|Y_{t},X_{t-1},\dots X_{0},Y_{t-1},\dots Y_{0})\\
&\times g_{t,Y}(Y_{t}|X_{t-1},\dots X_{0},Y_{t-1},\dots Y_{0})
\end{align*}
with
$$
g_{t,Y}(Y_{t}|X_{t-1},\dots X_{0},Y_{t-1},\dots Y_{0})=f_{Y|X}(Y_{t}|X_{t-1})
$$
which only depends on $X_{t-1},\dots X_{0},Y_{t-1},\dots Y_{0}$ through
$X_{t-1}$, according to Step 1. of Algorithm \ref{al:TSGibbs}. Moreover,
$$
g_{t,X|Y}(X_{t}|Y_{t},X_{t-1},\dots X_{0},Y_{t-1},\dots Y_{0})=f_{X|Y}(X_{t}|Y_{t})
$$
only depends on $X_{t-2},\dots X_{0},Y_{t},\dots Y_{0}$ through $Y_{t}$.
Therefore, 
$$
g_{t}(X_{t},Y_{t}|X_{t-1},\dots X_{0},Y_{t-1},\dots Y_{0})=g_{t}(X_{t},Y_{t}|X_{t-1})\,,
$$
which shows this is truly an homogeneous Markov chain.

\subsection{Exercise \ref{pb:multiAR}}

\begin{enumerate}
\renewcommand{\theenumi}{\alph{enumi}}
\item The (normal) full conditionals are defined in Example
\ref{ex:normgibbs2}. An \R program that implements this Gibbs
sampler is
\begin{verbatim}
# (C.) Anne Sabourin, 2009
T=500 ;p=5 ;r=0.25
X=cur=rnorm(p)
for (t in 1 :T){
  for (j in 1 :p){
    m=sum(cur[-j])/(p-1)
    cur[j]=rnorm(1,(p-1)*r*m/(1+(p-2)*r),
       sqrt((1+(p-2)*r-(p-1)*r^2)/(1+(p-2)*r)))
    }
  X=cbind(X,cur)
  }
par(mfrow=c(1,5))
for (i in 1:p){
   hist(X[i,],prob=TRUE,col="wheat2",xlab="",main="")
   curve(dnorm(x),add=TRUE,col="sienna",lwd=2)}
\end{verbatim}

\item Using instead
\begin{verbatim}
J=matrix(1,ncol=5,nrow=5)
I=diag(c(1,1,1,1,1))
s=(1-r)*I+r*J
rmnorm(500,s)
\end{verbatim}
and checking the duration by \verb+system.time+ shows \verb=rmnorm= is about five times
faster (and exact!).
\item If we consider the constraint
$$
\sum_{i=1}^{m} x_i^2 \le \sum_{i=m+1}^{p} x_i^2
$$
it imposes a truncated normal full conditional on {\em all} components. Indeed, for $1\le i\le m$,
$$
x^2_i \le \sum_{j=m+1}^{p} x_j^2 - \sum_{j=1,j\ne i}^{m} x_j^2\,,
$$
while, for $i>m$,
$$
x^2_i \ge \sum_{j=m+1,j\ne i}^{p} x_j^2 - \sum_{j=1}^{m} x_j^2\,.
$$
Note that the upper bound on $x_i^2$ when $i\le m$ {\em cannot be negative} if we start the Markov chain under the constraint.
The \verb#cur[j]=rnorm(...# line in the above \R program thus needs to be modified into a truncated normal distribution. 
An alternative is to use a hybrid solution (see Section \ref{sec:MwithinG} for the validation): 
we keep generating the $x_i$'s from the same plain normal full conditionals as before and we only
change the components for which the constraint remains valid, i.e.
\begin{verbatim}
  for (j in 1:m){
    mea=sum(cur[-j])/(p-1)
    prop=rnorm(1,(p-1)*r*mea/(1+(p-2)*r),
       sqrt((1+(p-2)*r-(p-1)*r^2)/(1+(p-2)*r)))
    if (sum(cur[(1:m)[-j]]^2+prop^2)<sum(cur[(m+1):p]^2))
      cur[j]=prop
    }
  for (j in (m+1):p){
    mea=sum(cur[-j])/(p-1)
    prop=rnorm(1,(p-1)*r*mea/(1+(p-2)*r),
       sqrt((1+(p-2)*r-(p-1)*r^2)/(1+(p-2)*r)))
    if (sum(cur[(1:m)]^2)<sum(cur[((m+1):p)[-j]]^2+prop^2))
      cur[j]=prop
    }
\end{verbatim}
Comparing the histograms with the normal $\mathcal{N}(0,1)$ shows that the marginals are no longer
normal.
\end{enumerate}

\subsection{Exercise \ref{pb:censoredGibbs}}

{\bf Warning: There is a typo in Example \ref{ex:censoredGibbs}, namely that the likelihood function involves
$\Phi(\theta-a)^{n-m}$ in front of the product of normal densities... For coherence with Examples 
\ref{ex:7.4.3.1}  and \ref{ex:EMCensored2}, in both Example \ref{ex:censoredGibbs} and Exercise \ref{pb:censoredGibbs},
$x$ should be written $\by$, $z$ $\bz$, $\bar x$ $\bar y$ and $x_i$ $y_i$.}

\begin{enumerate}
\renewcommand{\theenumi}{\alph{enumi}}
\item  The complete data likelihood is associated with the distribution of the uncensored data 
$$
(y_1,\ldots,y_m,z_{m+1},\ldots,z_n)\,,
$$
which constitutes an iid sample of size $n$. In that case, a sufficient statistics is $\{m\bar y+
(n-m(\bar z)\}/n$, which is distributed as $\mathcal{N}(\theta,1/n)$, i.e.~associated with the likelihood
$$
\exp\left\{ \dfrac{-n}{2}\,\left( \dfrac{m \bar x +(n-m) \bar z}{n} - \theta \right)^2 \right\}/\sqrt{n}\,.
$$
In this sense, the likelihood is proportional to the density of $\theta\sim{\mathcal N}(\{m \bar x +(n-m) \bar z\}/n,1/n )$.
(We acknowledge a certain vagueness in the wording of this question!)

\item The full \R code for the Gibbs sampler is
\begin{verbatim}
xdata=c(3.64,2.78,2.91,2.85,2.54,2.62,3.16,2.21,4.05,2.19,
2.97,4.32,3.56,3.39,3.59,4.13,4.21,1.68,3.88,4.33)
m=length(xdata)
n=30;a=3.5            #1/3 missing data
nsim=10^4
xbar=mean(xdata)
that=array(xbar,dim=c(nsim,1))
zbar=array(a,dim=c(nsim,1))
for (i in 2:nsim){       
   temp=runif(n-m,min=pnorm(a,mean=that[i-1],sd=1),max=1)
   zbar[i]=mean(qnorm(temp,mean=that[i-1],sd=1))
   that[i]=rnorm(1,mean=(m*xbar+(n-m)*zbar[i])/n,
                 sd=sqrt(1/n))
   }
par(mfrow=c(1,2),mar=c(5,5,2,1))
hist(that[500:nsim],col="grey",breaks=25,
xlab=expression(theta),main="",freq=FALSE)
curve(dnorm(x,mean(that),sd=sd(that)),add=T,lwd=2)
hist(zbar[500:nsim],col="grey",breaks=25
main="",xlab= expression(bar(Z)),freq=FALSE)
curve(dnorm(x,mean(zbar),sd=sd(zbar)),add=T,lwd=2)
\end{verbatim}
(We added the normal density curves to check how close to a normal distribution the posteriors are.)
\end{enumerate}

\subsection{Exercise \ref{pb:blood}}

\begin{enumerate}
\renewcommand{\theenumi}{\alph{enumi}}
\item Given the information provided in Table \ref{tab:GibbsBlood}, since we can reasonably assume independence
between the individuals, the distribution of the blood groups is a multinomial distribution whose density is
clearly proportional to 
$$
(p_A^2+2p_A p_O)^{n_A} (p_B^2+2p_B p_O)^{n_B}(p_A p_B)^{n_{AB}}(p_O^2)^{n_O}\,.
$$
the proportionality coefficient being the multinomial coefficient
$$
\left( \begin{matrix} &\ n& & \\n_A &n_B &n_{AB} &n_O\end{matrix} \right)\,.
$$
\item If we break $n_A$ into $Z_A$ individuals with genotype \verb+AA+ and $n_A-Z_A$ with genotype \verb+AO+, and
similarly, $n_B$ into $Z_B$ individuals with genotype \verb+BB+ and $n_B-Z_B$ with genotype \verb+BO+, the complete
data likelihood corresponds to the extended multinomial model with likelihood proportional to
$$
(p_A^2)^{Z_A}(2p_A p_O)^{n_A-Z_A} (p_B^2)^{Z_B}(2p_B p_O)^{n_B-Z_B}(p_A p_B)^{n_{AB}}(p_O^2)^{n_O}\,.
$$
\item The Gibbs sampler we used to estimate this model is
\begin{verbatim}
nsim=5000;nA=186;nB=38;nAB=13;nO=284;
pA=array(.25,dim=c(nsim,1));pB=array(.05,dim=c(nsim,1));
for (i in 2:nsim){
  pO=1-pA[i-1]-pB[i-1]
  ZA=rbinom(1,nA,pA[i-1]^2/(pA[i-1]^2+2*pA[i-1]*pO));
  ZB=rbinom(1,nB,pB[i-1]^2/(pB[i-1]^2+2*pB[i-1]*pO));
  temp=rdirichlet(1,c(nA+nAB+ZA+1,nB+nAB+ZB+1,
                  nA-ZA+nB-ZB+2*nO+1));
  pA[i]=temp[1];pB[i]=temp[2];
  }
par(mfrow=c(1,3),mar=c(4,4,2,1))
hist(pA,main=expression(p[A]),freq=F,col="wheat2")
hist(pB,main=expression(p[B]),freq=F,col="wheat2")
hist(1-pA-pB,,main=expression(p[O]),freq=F,col="wheat2")
\end{verbatim}
It uses the Dirichlet generator \verb+rdirichlet+ found in the \verb+mcsm+ package.
\end{enumerate}

\subsection{Exercise \ref{pb:slice}}

\begin{enumerate}
\renewcommand{\theenumi}{\alph{enumi}}
\item For the target density $f_{X}(x)=\frac{1}{2}e^{-\sqrt{x}}$, a slice sampling algorithm is
based on the full conditionals
\begin{enumerate}
\item $U^{(t+1)}\sim\mathcal{U}_{[0,f_{X}(x^{(t)})]}$
\item $X^{(t+1)}\sim\mathcal{U}_{A^{(t+1)}}$ with $A^{(t+1)}=\{ x,f(x)\geq u^{(t+1)}\}$
\end{enumerate}
Therefore, $U|x\sim\mathcal{U}(0,\frac{1}{2}e^{-\sqrt{x}})$ and, since
$A=\{ x,\frac{1}{2}e^{-\sqrt{x}}\geq u\}$,
i.e.~$A=\{ x,0\leq x\leq\log(2u)²\}$, owe also deduce that $X|u\sim\mathcal{U}(0,(\log(2u))^2)$.
The corresponding \R code is
\begin{verbatim}
T=5000
f=function(x){
   1/2*exp(-sqrt(x))}
X=c(runif(1)) ;U=c(runif(1))
for (t in 1:T){
  U=c(U,runif(1,0,f(X[t])))
  X=c(X,runif(1,0,(log(2*U[t+1]))^2))
  }
par(mfrow=c(1,2))
hist(X,prob=TRUE,col="wheat2",xlab="",main="")
acf(X)
\end{verbatim}
 
\item If we define $Y=\sqrt{X}$, then
\begin{align*}
P(Y \leq y) &= P(X\leq y^{2})\\
 &=\int_{0}^{y²}\frac{1}{2}e^{-\sqrt{x}}d\,\text{d}x
\end{align*}
When we differentiate against $y$, we get the density
$$
f_{Y}(y)=y\exp(-y)
$$
which implies that $Y\sim\mathcal{G}a(2,1)$.
Simulating $X$ then follows from $X=Y^2$.
This method is obviously faster and more accurate since the sample points
are then independent.
\end{enumerate}

\subsection{Exercise \ref{pb:normacf}}

\begin{enumerate}
\renewcommand{\theenumi}{\alph{enumi}}
\item The linear combinations $X+Y$ and $X-Y$ also are normal with null expectation and with variances
$2(1+\rho)$ and $2(1-\rho)$, respectively. The vector $(X+Y,X-Y)$ itself is equally normal.
Moreover,
$$
\text{cov}(X+Y,X-Y)=\mathbb{E}((X+Y)(X-Y))=\mathbb{E}(X^{2}-Y^{2})=1-1=0
$$
implies that $X+Y$ and $X-Y$ are independent.

\item If, instead, 
$$
(X,Y)\sim\mathcal{N}(0,\left(\begin{array}{cc}
\sigma_{x}^{2} & \rho\sigma_{x}\sigma_{y}\\
\rho\sigma_{x}\sigma_{y} & \sigma_{y}^{2}\end{array}\right))
$$
then $\sigma_{x}^{2}\neq\sigma_{y}^{2}$ implies that
$(X+Y)$ and $(X-Y)$ are dependent since $\mathbb{E}((X+Y)(X-Y))=\sigma_{x}^{2}-\sigma_{y}^{2}$.
In this case, $X|Y=y\sim\mathcal{N}(\rho\frac{\sigma_{x}}{\sigma_{y}}y,\sigma_{x}^{2}(1-\rho^{2}))$.
We can simulate $(X,Y)$ by the following Gibbs algorithm
\begin{verbatim}
T=5000;r=0.8;sx=50;sy=100
X=rnorm(1);Y=rnorm(1)
for (t in 1:T){
    Yn=rnorm(1,r*sqrt(sy/sx)*X[t],sqrt(sy*(1-r^2)))
    Xn=rnorm(1,r*sqrt(sx/sy)*Yn,sqrt(sx*(1-r^2)))
    X=c(X,Xn)
    Y=c(Y,Yn) 
    }
par(mfrow=c(3,2),oma=c(0,0,5,0))
hist(X,prob=TRUE,main="",col="wheat2")
hist(Y,prob=TRUE,main="",col="wheat2")
acf(X);acf(Y);plot(X,Y);plot(X+Y,X-Y)
\end{verbatim}
 
\item If $\sigma_{x}\neq\sigma_{y}$, let us find $a\in\mathbb{R}$ such that $X+aY$ and $Y$ are independent.
We have $\mathbb{E}[(X+aY)(Y)]=0$ if and only if $\rho\sigma_{x}\sigma_{y}+a\sigma_{y}^{2}=0$, 
i.e.~$a=-\rho\sigma_{x}/\sigma_{y}$. Therefore, $X-\rho\sigma_{x}/\sigma_{y}Y$ and $Y$ are independent.
\end{enumerate}

\subsection{Exercise \ref{pb:7.1}}  

\begin{enumerate}
\renewcommand{\theenumi}{\alph{enumi}}
\item  The likelihood function naturally involves the tail of the Poisson distribution
for those observations larger than $4$. The full conditional distributions of the observations larger than $4$
are obviously truncated Poisson distributions and the full conditional distribution of the parameter is the
Gamma distribution associated with a standard Poisson sample. Hence the Gibbs sampler.
\item The \R code we used to produce Figure \ref{fig:PoissonRB} is
\begin{verbatim}
nsim=10^3
lam=RB=rep(313/360,nsim)
z=rep(0,13)
for (j in 2:nsim){
  top=round(lam[j -1]+6*sqrt(lam[j -1]))
  prob=dpois(c(4:top),lam[j -1])
  cprob=cumsum(prob/sum(prob))
  for(i in 1:13) z[i] = 4+sum(cprob<runif(1))
  RB[j]=(313+sum(z))/360
  lam[j]=rgamma(1,360*RB[j],scale=1/360);
  }
par(mfrow=c(1,3),mar=c(4,4,2,1))
hist(lam,col="grey",breaks=25,xlab="",
     main="Empirical average")
plot(cumsum(lam)/1:nsim,ylim=c(1,1.05),type="l",
     lwd=1.5,ylab="")
lines(cumsum(RB)/1:nsim,col="sienna",lwd=1.5)
hist(RB,col="sienna",breaks=62,xlab="",
     main="Rao-Blackwell",xlim=c(1,1.05))
\end{verbatim}
\item When checking the execution time of both programs with \verb+system.time+, 
the first one is almost ten times faster. And completely correct. A natural way
to pick \verb+prob+ is
\begin{verbatim}
> qpois(.9999,lam[j-1])
[1] 6
\end{verbatim}
\end{enumerate}

\subsection{Exercise \ref{pb:Exp-Improper}}  

\begin{enumerate}
\renewcommand{\theenumi}{\alph{enumi}}
\item  The \R program that produced Figure \ref{fig:Exp-Improper} is
\begin{verbatim}
nsim=10^3
X=Y=rep(0,nsim)
X[1]=rexp(1)            #initialize the chain
Y[1]=rexp(1)            #initialize the chain
for(i in 2:nsim){
        X[i]=rexp(1,rate=Y[i-1])
        Y[i]=rexp(1,rate=X[i])
        }
st=0.1*nsim
par(mfrow=c(1,2),mar=c(4,4,2,1))
hist(X,col="grey",breaks=25,xlab="",main="")
plot(cumsum(X)[(st+1):nsim]/(1:(nsim-st)),type="l",ylab="")
\end{verbatim}
\item Using the Hammersley--Clifford Theorem {\em per se} means using $f(y|x)/f(x|y)=x/y$ which is {\em not integrable}.
If we omit this major problem, we have
$$
f(x,y) = \frac{x\,\exp\{-xy\}}{x\, {\displaystyle \int \dfrac{\text{d}y}{y}}} \propto \exp\{-xy\}
$$
(except that the proportionality term is infinity!).
\item If we constrain both conditionals to $(0,B)$, the Hammersley--Clifford Theorem gives
\begin{align*}
f(x,y) &= \frac{\exp\{-xy\}/(1-e^{-xB})}{{\displaystyle \int \dfrac{1-e^{-yB}}{y(1-e^{-xB})}\,\text{d}y}}\\
       &= \frac{\exp\{-xy\}}{{\displaystyle \int \dfrac{1-e^{-yB}}{y}\,\text{d}y}}\\
       &\propto \exp\{-xy\}\,,
\end{align*}
since the conditional exponential distributions are truncated. This joint distribution is then well-defined on
$(0,B)^2$. A Gibbs sampler simulating from this joint distribution is for instance
\begin{verbatim}
B=10
X=Y=rep(0,nsim)
X[1]=rexp(1)            #initialize the chain
Y[1]=rexp(1)            #initialize the chain
for(i in 2:nsim){	#inversion method
        X[i]=-log(1-runif(1)*(1-exp(-B*Y[i-1])))/Y[i-1]
        Y[i]=-log(1-runif(1)*(1-exp(-B*X[i])))/X[i]
        }
st=0.1*nsim
marge=function(x){ (1-exp(-B*x))/x}
nmarge=function(x){ 
       marge(x)/integrate(marge,low=0,up=B)$val}
par(mfrow=c(1,2),mar=c(4,4,2,1))
hist(X,col="wheat2",breaks=25,xlab="",main="",prob=TRUE)
curve(nmarge,add=T,lwd=2,col="sienna")
plot(cumsum(X)[(st+1):nsim]/c(1:(nsim-st)),type="l",
     lwd=1.5,ylab="")
\end{verbatim}
where the simulation of the truncated exponential is done by inverting the cdf (and where the
true marginal is represented against the histogram).
\end{enumerate}

\subsection{Exercise \ref{pb:firsthier}}

Let us define
\begin{eqnarray*}
f(x) & = & \frac{b^{a}x^{a-1}e^{-bx}}{\Gamma(a)}\,,\\
g(x) & = & \frac{1}{x}=y\,,\end{eqnarray*}
then we have
\begin{eqnarray*}
f_{Y}(y) & = & f_{X}\left(g^{-1}(y)\right)\mid\frac{d}{dy}g^{-1}(y)\mid\\
 & = & \frac{b^{a}}{\Gamma(a)}\left({1}/{y}\right)^{a-1}\exp\left(-{b}/{y}\right)\frac{1}{y^{2}}\\
 & = & \frac{b^{a}}{\Gamma(a)}\left({1}/{y}\right)^{a+1}\exp\left(-{b}/{y}\right)\,,
\end{eqnarray*}
which is the ${\cal IG}(a,b)$ density.

\subsection{Exercise \ref{pb:truncnorm}}

{\bf Warning: The function \verb+rtnorm+ requires a predefined \verb+sigma+ that should be part
of the arguments, as in\\ 
\verb+rtnorm=function(n=1,mu=0,lo=-Inf,up=Inf,sigma=1)+.}\\

Since the \verb+rtnorm+ function is exact (within the precision of the \verb+qnorm+ and \verb+pnorm+
functions, the implementation in \R is straightforward:
\begin{verbatim}
h1=rtnorm(10^4,lo=-1,up=1)
h2=rtnorm(10^4,up=1)
h3=rtnorm(10^4,lo=3)
par(mfrow=c(1,3),mar=c(4,4,2,1))
hist(h1,freq=FALSE,xlab="x",xlim=c(-1,1),col="wheat2")
dnormt=function(x){ dnorm(x)/(pnorm(1)-pnorm(-1))}
curve(dnormt,add=T,col="sienna")
hist(h2,freq=FALSE,xlab="x",xlim=c(-4,1),col="wheat2")
dnormt=function(x){ dnorm(x)/pnorm(1)}
curve(dnormt,add=T,col="sienna")
hist(h3,freq=FALSE,xlab="x",xlim=c(3,5),col="wheat2")
dnormt=function(x){ dnorm(x)/pnorm(-3)}
curve(dnormt,add=T,col="sienna")
\end{verbatim}

\subsection{Exercise \ref{pb:freq_2}}

\begin{enumerate}
\renewcommand{\theenumi}{\alph{enumi}}
\item Since $(j=1,2)$
$$
(1-\theta_1-\theta_2)^{x_5+\alpha_3-1} = \sum_{i=0}^{x_5+\alpha_3-1}
{x_5+\alpha_3-1\choose i} (1-\theta_j)^i\theta_{3-j}^{x_5+\alpha_3-1-i}\,,
$$
when $\alpha_3$ is an integer, it is clearly possible to express $\pi(\theta_1,\theta_2|x)$ as
a sum of terms that are products of a polynomial function of $\theta_1$ and of a polynomial 
function of $\theta_2$.  It is therefore straightforward to integrate those terms in either $\theta_1$ 
or $\theta_2$.
\item For the same reason as above, rewriting $\pi(\theta_1,\theta_2|x)$ as a density in $(\theta_1,\xi)$
leads to a product of polynomials in $\theta_1$, all of which can be expanded and integrated in $\theta_1$,
producing in the end a sum of functions of the form 
$$
\xi^{\delta}\big/(1+\xi)^{x_1+x_2+x_5+\alpha_1+\alpha_3-2}\,,
$$
namely a mixture of $F$ densities.
\item The Gibbs sampler based on (\ref{eq:tannerFull}) is available in the \verb+mcsm+ package.
\end{enumerate}

\subsection{Exercise \ref{pb:RBall}} 

{\bf Warning: There is a typo in Example 7.3, \verb+sigma+ should be defined as \verb+sigma2+
and \verb+sigma2{1}+ should be \verb+sigma2[1]+...}\\

\begin{figure}
\centerline{\includegraphics[width=\textwidth]{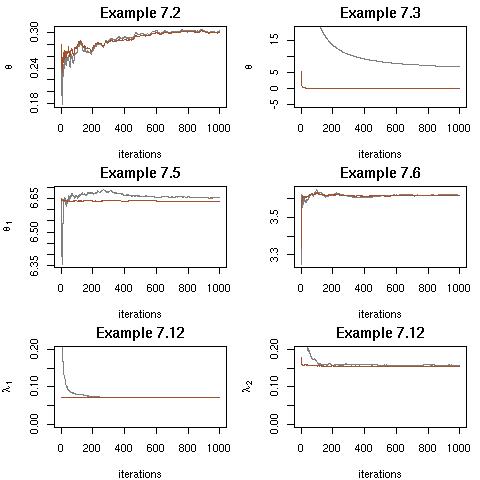}}
\caption{\label{fig:allrb's}
Comparison of the convergences of the plain average with its Rao-Blackwellized counterpart for
five different examples. The Rao-Blackwellized is plotted in {\sf sienna} red and is always more
stable than the original version.}
\end{figure}

\begin{enumerate}
\renewcommand{\theenumi}{\alph{enumi}}
\item  In Example \ref{ex:betabi}, since $\theta|x\sim {\cal B}e(x+a,n-x+b)$, we have clearly  $\BE[\theta \vert x] = (x+a)/(n+a+b)$ (with a missing
parenthesis). The comparison between the empirical average and of the Rao--Blackwellization version is of the form
\begin{verbatim}
plot(cumsum(T)/(1:Nsim),type="l",col="grey50",
     xlab="iterations",ylab="",main="Example 7.2")
lines(cumsum((X+a))/((1:Nsim)*(n+a+b)),col="sienna")
\end{verbatim}
All comparisons are gathered in Figure \ref{fig:allrb's}.

\item  In Example \ref{ex:Metab-1}, equation (\ref{eq:firstposterior}) defines two standard distributions as full
conditionals. Since $\pi(\theta|\bx,\sigma^2)$ is a normal distribution with mean and variance provided two lines
below, we obviously have  
$$
\BE[\theta | \bx,\sigma^2] = \frac{\sigma^2}{\sigma^2+n \tau^2}\;\theta_0 + \frac{n\tau^2}{\sigma^2+n \tau^2} \;\bar x
$$
The modification in the \R program follows
\begin{verbatim}
plot(cumsum(theta)/(1:Nsim),type="l",col="grey50",
     xlab="iterations",ylab="",main="Example 7.3")
ylab="",main="Example 7.3")
lines(cumsum(B*theta0+(1-B)*xbar)/(1:Nsim)),col="sienna")
\end{verbatim}

\item The full conditionals of Example \ref{ex:Metab-2} given in Equation (\ref{eq:onewayfull})
are more numerous but similarly standard, therefore 
$$
\BE[\theta_i | \bar X_i ,\sigma^2] = \frac{\sigma^2 }{\sigma^2+n_i \tau^2} \mu+\frac{n_i \tau^2 }{\sigma^2+n_i \tau^2}\bar X_i
$$
follows from this decomposition, with the \R lines added to the \verb+mcsm+ \verb+randomeff+ function
\begin{verbatim}
plot(cumsum(theta1)/(1:nsim),type="l",col="grey50",
     xlab="iterations",ylab="",main="Example 7.5")
lines(cumsum((mu*sigma2+n1*tau2*x1bar)/(sigma2+n1*tau2))/
      (1:nsim)),col="sienna")
\end{verbatim}

\item  In Example \ref{ex:censoredGibbs}, the complete-data model is a standard normal model with
variance one, hence  $\BE[\theta \vert x, z ] = \dfrac{m \bar x +(n-m) \bar z}{n}$. The additional lines
in the \R code are
\begin{verbatim}
plot(cumsum(that)/(1:Nsim),type="l",col="grey50",
     xlab="iterations",ylab="",main="Example 7.6")
lines(cumsum((m/n)*xbar+(1-m/n)*zbar)/(1:Nsim)),
      col="sienna")
\end{verbatim}

\item  In Example \ref{ex:5.7}, the full conditional on $\lambda$,
$\lambda_i|\beta,t_i,x_i \sim \CG (x_i+\alpha,t_i+\beta)$ and hence 
$\BE[\lambda_i|\beta,t_i,x_i] =  (x_i+\alpha)/(t_i+\beta)$. The corresponding addition
in the \R code is
\begin{verbatim}
plot(cumsum(lambda[,1])/(1:Nsim),type="l",col="grey50",
     xlab="iterations",ylab="",main="Example 7.12")
lines(cumsum((xdata[1]+alpha)/(Time[1]+beta))/(1:Nsim)),
      col="sienna")
\end{verbatim}
\end{enumerate}

\chapter{Convergence Monitoring for MCMC Algorithms}

\subsection{Exercise \ref{exo:lem:6.1}} 

{\bf Warning: Strictly speaking, we need to assume that the Markov chain $(x^{(t)})$ has a
finite variance for the $h$ transform, since the assumption that $\mathbb{E}_f[h^2(X)]$ exists is not
sufficient (see \citealp{meyn:tweedie:1993}.}

This result was established by \cite{maceachern:berliner:1994}.
We have the proof detailed as Lemma 12.2 in \cite{robert:casella:2004} (with the same
additional assumption on the convergence of the Markov chain missing!). 

Define $\delta_k^1,\ldots,\delta_k^{k-1}$
as the shifted versions of $\delta_k = \delta_k^0$; that is,
$$
\delta_k^i = {1 \over T} \; \sum_{t=1}^{T} \; h(\theta^{(tk-i)}) ,
\qquad\qquad i=0,1,\ldots,k-1 \;.
$$
The estimator $\delta_1$ can then be written as $\delta_1 =
{1 \over k} \; \sum_{i=0}^{k-1} \; \delta_k^i $, and hence
\begin{eqnarray*}
{\mathrm {var}}(\delta_1) &=& \displaystyle{ {\mathrm {var}}\left({1 \over k} \;
\sum_{i=0}^{k-1} \; \delta_k^i \right) } \\
&=& \displaystyle{ {\mathrm {var}}(\delta_k^0)/k + \sum_{i\neq j} \;
{\mathrm {cov}}(\delta_k^i,\delta_k^j) / k^2 } \\
&\leq& \displaystyle{ {\mathrm {var}}(\delta_k^0)/k + \sum_{i\neq j} \;
{\mathrm {var}}(\delta_k^0) / k^2 } \\
&=& \displaystyle{ {\mathrm {var}}(\delta_k)\;, }
\end{eqnarray*}
where the inequality follows from the Cauchy--Schwarz inequality
$$
|\text{cov}(\delta_k^i, \delta_k^j)| \leq \text{var}(\delta_k^0).
$$
 
\subsection{Exercise \ref{exo:misgim}}

This is a direct application of the Ergodic Theorem (see Section \ref{sec:dumdum}).
If the chain $(x^{(t)})$ is ergodic, then the empirical average above converges (almost
surely) to $\mathbb{E}_f[\varphi(X) \big/ \tilde f(X)]=1/C$. This assumes that the support of
$\varphi$ is {\em small enough} (see Exercise \ref{pb:ratio_csts3}). For the variance of the
estimator to be finite, a necessary condition is that
$$
\mathbb{E}_f[\varphi(X) \big/ \tilde f(X)] \propto \int \dfrac{\varphi^2(x)}{f(x)}\,\text{d}x < \infty\,.
$$
As in Exercise \ref{exo:lem:6.1}, we need to assume that the convergence of the Markov chain
is regular enough to ensure a finite variance.

\subsection{Exercise \ref{exo:patchtrap}}

\begin{figure}
\centerline{\includegraphics[width=\textwidth]{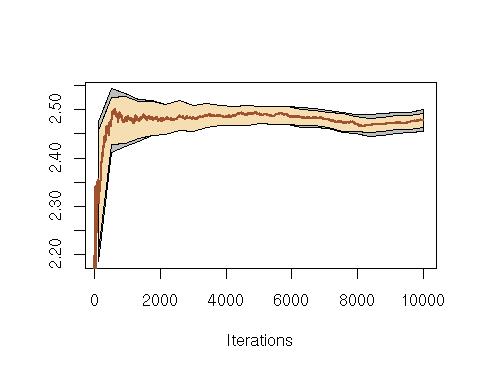}}
\caption{\label{fig:bootband}
Comparison of two evaluations of the variance of the MCMC estimate of the mean of $\beta$ for
the pump failure model of Example \ref{ex:patchump}.}
\end{figure}

The modified \R program using bootstrap is
\begin{verbatim}
ranoo=matrix(0,ncol=2,nrow=25)
for (j in 1:25){
 batch=matrix(sample(beta,100*Ts[j],rep=TRUE),ncol=100)
 sigmoo=2*sd(apply(batch,2,mean)) 
 ranoo[j,]=mean(beta[1:Ts[j]])+c(-sigmoo,+sigmoo)
 }
polygon(c(Ts,rev(Ts)),c(ranoo[,1],rev(ranoo[,2])),col="grey")
lines(cumsum(beta)/(1:T),col="sienna",lwd=2)
\end{verbatim}
and the output of the comparison is provided in Figure \ref{fig:bootband}.

\subsection{Exercise \ref{exo:baseball}}

{\bf Warning: Example \ref{ex:baseball} contains several typos, namely $Y_k\sim\CN(\theta_i,\sigma^2)$
instead of $Y_i\sim\CN(\theta_i,\sigma^2)$, {\sf the $\mu_i$'s being also
iid normal} instead of {\sf the $\theta_i$'s being also iid normal}...}\\

{\bf Warning: Exercise \ref{exo:baseball} also contains a typo in that the posterior distribution on $\mu$ 
cannot be obtained in a closed form. It should read}
\begin{rema}\noindent
Show that the posterior distribution on $\alpha$ in Example \ref{ex:baseball} can be obtained in a closed form.
\end{rema}

Since 
\begin{align*}
\mathbf{\theta} | \by,\mu,\alpha &\sim \pi(\mathbf{\theta}|\by,\mu,\alpha)\\
 &\propto \alpha^{-9} \exp\dfrac{-1}{2}\left\{ \sum_{i=1}^{18} \left[
\sigma^{-2} (y_i-\theta_i)^2 + \alpha^{-1}(\theta_i-\mu)^2 \right] \right\}\\
&\propto \exp\dfrac{-1}{2}\left(\sum_{i=1}^{18} \left\{
(\sigma^{-2} + \alpha^{-1})\left[\theta_i-(\sigma^{-2} + \alpha^{-1})^{-1}(\sigma^{-2} y_i 
+\alpha^{-1} \mu)\right]^2\right.\right.\\
&\left.\left.\quad + (\alpha+\sigma^2)^{-1}\sum_{i=1}^{18} (y_i-\mu)^2 \right\}\right)
\end{align*}
(which is also a direct consequence of the marginalization $Y_i\sim\CN(\mu,\alpha+\sigma^2)$), we have
\begin{align*}
\pi(\alpha,\mu|\by) &\propto \dfrac{\alpha^{-3}}{(\alpha+\sigma^2)^{9}}\, \exp\left\{-\dfrac{1}{2(\alpha+\sigma^2)}
\sum_{i=1}^{18} (y_i-\mu)^2 -\dfrac{\mu^2}{2}-\dfrac{2}{\alpha} \right\}\\
&\propto \dfrac{\alpha^{-3}}{(\alpha+\sigma^2)^{9}}\, \exp\bigg\{-\dfrac{2}{\alpha} \\
&\quad-\dfrac{1+n(\alpha+\sigma^2)^{-1}}{2}\left[
\mu-(\alpha+\sigma^2)^{-1}\sum_{i=1}^{18} y_i\big/(1+n(\alpha+\sigma^2)^{-1})\right]^2\\
&\quad\left.-\dfrac{1}{2(\alpha+\sigma^2)}\sum_{i=1}^{18} y_i^2 + \dfrac{(\alpha+\sigma^2)^{-2}}{2(1+n(\alpha+\sigma^2)^{-1})}
\left(\sum_{i=1}^{18} y_i\right)^2 \right\}
\end{align*}
and thus
\begin{align*}
\pi(\alpha|\by) 
&\propto \dfrac{\alpha^{-3}(1+n(\alpha+\sigma^2)^{-1})^{-1/2}}{(\alpha+\sigma^2)^{9}}\, 
\exp\bigg\{-\dfrac{2}{\alpha} \\
&\quad\left.-\dfrac{1}{\alpha+\sigma^2}\sum_{i=1}^{18} y_i^2 + \dfrac{(\alpha+\sigma^2)^{-2}}{1+n(\alpha+\sigma^2)^{-1}}
\left(\sum_{i=1}^{18} y_i\right)^2 \right\}
\end{align*}
Therefore the marginal posterior distribution on $\alpha$ has a closed (albeit complex) form. (It is also
obvious from $\pi(\alpha,\mu|\by)$ above that the marginal posterior on $\mu$ does not have a closed form.)

The baseball dataset can be found in the \verb+amcmc+ package in the \verb+baseball.c+ program and rewritten as
\begin{verbatim}
baseball=c(0.395,0.375,0.355,0.334,0.313,0.313,0.291,
0.269,0.247,0.247,0.224,0.224,0.224,0.224,0.224,0.200,
0.175,0.148)
\end{verbatim}

The standard Gibbs sampler is implemented by simulating
\begin{align*}
\theta_i|y_i,\mu,\alpha &\sim \mathcal{N}\left(\dfrac{\alpha^{-1}\mu+\sigma^{-2}y_i}{\alpha^{-1}+\sigma^{-2}},
(\alpha^{-1}+\sigma^{-2})^{-1} \right)\,,\\
\mu|\mathbf{\theta},\alpha &\sim \mathcal{N}\left(\dfrac{\alpha^{-1}\sum_{i=1}^{18}\theta_i}{1+n\alpha^{-1}},
(n\alpha^{-1}+1)^{-1} \right)\,,\\
\alpha|\mathbf{\theta},\mu&\sim\mathcal{IG}\left(11,2+\sum_{i=1}^{18} (\theta_i-\mu)^2/2 \right)
\end{align*}
which means using an \R loop like
\begin{verbatim}
Nsim=10^4
sigma2=0.00434;sigmam=1/sigma2
theta=rnorm(18)
mu=rep(rnorm(1),Nsim)
alpha=rep(rexp(1),Nsim)
for (t in 2:Nsim){
  theta=rnorm(18,mean=(mu[t-1]/alpha[t-1]+sigmam*baseball)/
  (1/alpha[t-1]+sigmam),sd=1/sqrt(1/alpha[t-1]+sigmam))
  mu[t]=rnorm(1,mean=sum(theta)/(1/alpha[t-1]+n),
        sd=1/sqrt(1+n/alpha[t-1]))
  alpha[t]=(2+0.5*sum((theta-mu[t])^2))/rgamma(1,11)
}
\end{verbatim}
The result of both \verb+coda+ diagnostics on $\alpha$ is
\begin{verbatim}
> heidel.diag(mcmc(alpha))

     Stationarity start     p-value
     test         iteration
var1 passed       1         0.261

     Halfwidth Mean  Halfwidth
     test
var1 passed    0.226 0.00163
> geweke.diag(mcmc(alpha))

Fraction in 1st window = 0.1
Fraction in 2nd window = 0.5

   var1
-0.7505
\end{verbatim}
If we reproduce the Kolmogorov--Smirnov analysis
\begin{verbatim}
ks=NULL
M=10
for (t in seq(Nsim/10,Nsim,le=100)){
alpha1=alpha[1:(t/2)]
alpha2=alpha[(t/2)+(1:(t/2))]
alpha1=alpha1[seq(1,t/2,by=M)]
alpha2=alpha2[seq(1,t/2,by=M)]
ks=c(ks,ks.test(alpha1,alpha2)$p)
}
\end{verbatim}
Plotting the vector \verb+ks+ by \verb+plot(ks,pch=19)+ 
shows no visible pattern that would indicate a lack of uniformity.

\begin{figure}
\centerline{\includegraphics[width=0.7\textwidth]{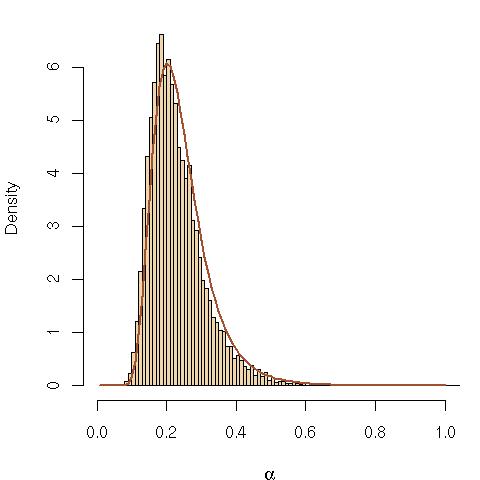}}
\caption{\label{fig:dafit}
Histogram of the $(\alpha^{(t)})$ chain produced by the Gibbs sampler of Example \ref{ex:baseball} 
and fit of the exact marginal $\pi(\alpha|\by)$, based on $10^4$ simulations.}
\end{figure}

Comparing the output with the true target in $\alpha$ follows from the definition
\begin{verbatim}
marge=function(alpha){
(alpha^(-3)/(sqrt(1+18*(alpha+sigma2)^(-1))*(alpha+sigma2)^9))*
exp(-(2/alpha) - (.5/(alpha+sigma2))*sum(baseball^2) +
.5*(alpha+sigma2)^(-2)*sum(baseball)^2/(1+n*(alpha+sigma2)^(-1)))
}
\end{verbatim}
Figure \ref{fig:dafit} shows the fit of the simulated histogram to the above function (when normalized
by \verb+integrate+).

\subsection{Exercise \ref{pb:the_far_side}}

\begin{enumerate}
 \renewcommand{\theenumi}{\alph{enumi}}
\item We simply need to check that this transition kernel $K$ satisfies the
detailed balance condition \eqref{eq:db}, $f(x)K(y|x) =  f(y) K(x|y)$ when $f$
is the ${\cal B}e(\alpha,1)$ density: when $x\ne y$,
\begin{align*}
f(x)K(x,y) &= \alpha x^{\alpha-1}\,x\,(\alpha+1)\,y^{\alpha}\\
	   &= \alpha (\alpha+1) (xy)^\alpha\\
           &= f(y)K(y,x)
\end{align*}
so the ${\cal B}e(\alpha,1)$ distribution is indeed stationary.
\item Simulating the Markov chain is straightforward:
\begin{verbatim}
alpha=.2
Nsim=10^4
x=rep(runif(1),Nsim)
y=rbeta(Nsim,alpha+1,1)
for (t in 2:Nsim){
   if (runif(1)<x[t-1]) x[t]=y[t]
   else	x[t]=x[t-1]
   } 
\end{verbatim}
and it exhibits a nice fit to the beta ${\cal B}e(\alpha,1)$ target. However,
running \verb+cumuplot+ shows a lack of concentration of the distribution, while
the two standard stationarity diagnoses are
\begin{verbatim}
> heidel.diag(mcmc(x))

     Stationarity start     p-value
     test         iteration
var1 passed       1001      0.169

     Halfwidth Mean  Halfwidth
     test
var1 failed    0.225 0.0366
> geweke.diag(mcmc(x))

Fraction in 1st window = 0.1
Fraction in 2nd window = 0.5

 var1
3.277
\end{verbatim}
are giving dissonant signals. The \verb+effectiveSize(mcmc(x))}+ is then equal to $329$.
Moving to $10^6$ simulations does not modify the picture (but may cause your system to crash!)
\item The corresponding Metropolis--Hastings version is
\begin{verbatim}
alpha=.2
Nsim=10^4
x=rep(runif(1),Nsim)
y=rbeta(Nsim,alpha+1,1)
for (t in 2:Nsim){
   if (runif(1)<x[t-1]/y[t]) x[t]=y[t]
   else x[t]=x[t-1]
   }
\end{verbatim}
It also provides a good fit and also fails the test:
\begin{verbatim}
> heidel.diag(mcmc(x))

     Stationarity start     p-value
     test         iteration
var1 passed       1001      0.0569

     Halfwidth Mean  Halfwidth
     test
var1 failed    0.204 0.0268
> geweke.diag(mcmc(x))

Fraction in 1st window = 0.1
Fraction in 2nd window = 0.5

 var1
1.736
\end{verbatim}
\end{enumerate}

\subsection{Exercise \ref{pb:proberge}}

\begin{enumerate}
\renewcommand{\theenumi}{\alph{enumi}}
\item A possible \R definition of the posterior is
\begin{verbatim}
postit=function(beta,sigma2){
  prod(pnorm(r[d==1]*beta/sigma2))*prod(pnorm(-r[d==0]*beta/sigma2))*
  dnorm(beta,sd=5)*dgamma(1/sigma2,2,1)}
\end{verbatim}
and a possible \R program is
\begin{verbatim}
r=Pima.tr$ped
d=as.numeric(Pima.tr$type)-1
mod=summary(glm(d~r-1,family="binomial"))
beta=rep(mod$coef[1],Nsim)
sigma2=rep(1/runif(1),Nsim)
for (t in 2:Nsim){
  prop=beta[t-1]+rnorm(1,sd=sqrt(sigma2[t-1]*mod$cov.unscaled))
  if (runif(1)<postit(prop,sigma2[t-1])/postit(beta[t-1],
	sigma2[t-1])) beta[t]=prop
  else beta[t]=beta[t-1]
  prop=exp(log(sigma2[t-1])+rnorm(1))
  if (runif(1)<sigma2[t-1]*postit(beta[t],prop)/(prop*
      postit(beta[t], sigma2[t-1]))) sigma2[t]=prop
  else sigma2[t]=sigma2[t-1]
  }
\end{verbatim}
(Note the Jacobian $1/\sigma^2$ in the acceptance probability.)
\item Running $5$ chains in parallel is easily programmed with an additional loop
in the above. Running \verb+gelman.diag+ on those five chains then produces a
convergence assessment:
\begin{verbatim}
> gelman.diag(mcmc.list(mcmc(beta1),mcmc(beta2),mcmc(beta3),
+ mcmc(beta4),mcmc(beta5)))
Potential scale reduction factors:
     Point est. 97.5% quantile
[1,]       1.02           1.03
\end{verbatim}
Note also the good mixing behavior of the chain:
\begin{verbatim}
> effectiveSize(mcmc.list(mcmc(beta1),mcmc(beta2),
+ mcmc(beta3),mcmc(beta4),mcmc(beta5)))
    var1
954.0543
\end{verbatim}
\item The implementation of the traditional Gibbs sampler with completion is
detailed in \cite{marin:robert:2007}, along with the appropriate \R program.
The only modification that is needed for this problem is the introduction of 
the non-identifiable scale factor $\sigma^2$.
\end{enumerate}

\subsection{Exercise \ref{pb:thin_ks}}

In the \verb+kscheck.R+ program available in \verb+mcsm+, you can modify $G$ by
changing the variable \verb+M+ in
\begin{verbatim}
subbeta=beta[seq(1,T,by=M)]
subold=oldbeta[seq(1,T,by=M)]
ks=NULL
for (t in seq((T/(10*M)),(T/M),le=100)) 
   ks=c(ks,ks.test(subbeta[1:t],subold[1:t])$p)
\end{verbatim}
(As noted by a reader, the syntax \verb+ks=c(ks,res)+ is very inefficient in
system time, as you can check by yourself.)

\subsection{Exercise \ref{pb:tan_cvg}}

Since the Markov chain $(\theta^{(t)})$ is converging to the posterior distribution
(in distribution), the density at time $t$, $\pi_t$, is also converging (pointwise)
to the posterior density $\pi(\theta|x)$, therefore $\omega_t$ is converging to
$$
\dfrac{f(x|\theta^{(\infty)}) \pi(\theta^{(\infty)})}{ \pi(\theta^{(\infty)}|x)} = m(x)\,,
$$
for all values of $\theta^{(\infty)}$. (This is connected with Chib's (\citeyear{chib:1995})
method, discussed in Exercise \ref{exo:chibmarge}.)

\subsection{Exercise \ref{pb:essPress}}

If we get back to Example \ref{ex:6.1}, the sequence \verb+beta+ can be checked in terms of
effective sample via an \R program like
\begin{verbatim}
ess=rep(1,T/10)
for (t in 1:(T/10)) ess[t]=effectiveSize(beta[1:(10*t)])
\end{verbatim}
where the subsampling is justified by the computational time required by \verb&effectiveSize&.
The same principle can be applied to any chain produced by an MCMC algorithm.

\begin{figure}
\centerline{\includegraphics[width=\textwidth]{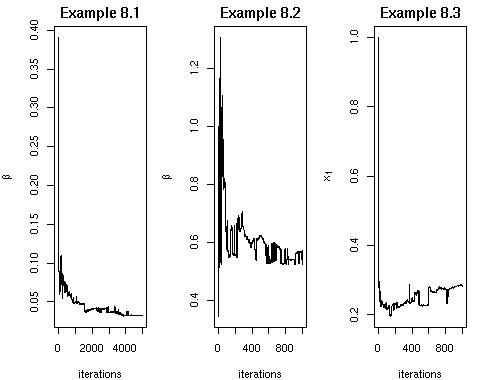}}
\caption{\label{esscomp}
Evolution of the effective sample size across iterations for the first three examples of
Chapter 8.}
\end{figure}

Figure \ref{esscomp} compares the results of this evaluation over the first three examples of
this chapter. None of them is strongly conclusive about convergence...


\backmatter


\begin{thebibliography}{}

\bibitem[Chib, 1995]{chib:1995}
Chib, S. (1995).
\newblock Marginal likelihood from the {G}ibbs output.
\newblock {\em Journal of the American Statistical Association}, 90:1313--1321.

\bibitem[MacEachern and Berliner, 1994]{maceachern:berliner:1994}
MacEachern, S. and Berliner, L. (1994).
\newblock Subsampling the {G}ibbs sampler.
\newblock {\em The American Statistician}, 48:188--190.

\bibitem[Marin and Robert, 2007]{marin:robert:2007}
Marin, J.-M. and Robert, C. (2007).
\newblock {\em Bayesian Core}.
\newblock Springer--Verlag, New York.

\bibitem[Meyn and Tweedie, 1993]{meyn:tweedie:1993}
Meyn, S. and Tweedie, R. (1993).
\newblock {\em {M}arkov Chains and Stochastic {S}tability}.
\newblock Springer--Verlag, New York.

\bibitem[Robert and Casella, 2004]{robert:casella:2004}
Robert, C. and Casella, G. (2004).
\newblock {\em {M}onte {C}arlo Statistical Methods, {\em second edition}}.
\newblock Springer--Verlag, New York.

\bibitem[Robert and Marin, 2010]{robert:marin:2010}
Robert, C. and Marin, J.-M. (2010).
\newblock Importance sampling methods for {B}ayesian discrimination between
  embedded models.
\newblock In Chen, M.-H., Dey, D.~K., Mueller, P., Sun, D., and Ye, K.,
  editors, {\em Frontiers of Statistical Decision Making and {B}ayesian
  Analysis}.
\newblock (To appear.).

\end{thebibliography}
\end{document}